Effect of microstructural inhomogeneity on load-dependent mechanical response and thermally induced shape memory recovery in plasma-sprayed NiTi coatings

**Jan Tomáštík[a,b*], Lukáš Václavek[a,b], Jaromír Kopeček[c], Akhil Bhardwaj[c], Tadeáš Těhan[d], Ondřej Tyc[c], Sneha Samal[c**]**

[a]Palacký University in Olomouc, Faculty of Science, Joint Laboratory of Optics of Palacký University and Institute of Physics AS CR, 17. listopadu 12, 779 00 Olomouc, Czech Republic

[b]Institute of Physics of the Czech Academy of Sciences, Joint Laboratory of Optics of Palacký University and Institute of Physics of the Czech Academy of Sciences, 779 00 Olomouc, Czech Republic

[c]Institute of Physics of the Czech Academy of Sciences, 182 21 Prague, Czech Republic

[d]Faculty of Electrical Engineering, Czech Technical University, Prague, Czech Republic

Corresponding Authors: *Tomáštík Jan, email: tomastik@fzu.cz, **Sneha Samal, email: samal@fzu.cz,

Highlights

- Recovery on unloading and on heating partitioned on the same indents in sprayed NiTi.
- Both elastic and thermal recovery decrease with load; 25–50 mN imprints heal almost fully.
- Depth and recovery scatter reflects the heterogeneity caused by the plasma-sprayed microstructure; they both decrease with higher indentation load
- The most porous deposit recovers least on both unloading and heating

Abstract

Plasma spraying can consolidate NiTi powder into thick, near-net-shape deposits, but the rapid melting and solidification of individual splats leave a heterogeneous microstructure of austenite, martensite, intermetallic and oxide phases with marked porosity. How this heterogeneity affects the ability of the surface to recover local contact damage has not yet been quantified. Here, four NiTi deposits sprayed at 9 and 12 kW with powder feed rates of 2.1 and 4.2 g/min were indented with a conospherical 10 µm tip at loads of 25–500 mN, and the residual imprints were profiled by confocal microscopy before and after annealing at 150 °C. Two types of recovery were quantified on the same indents: elastic recovery after unloading, from the load–displacement curves, and thermal recovery of the residual depth upon heating. Both decreased with increasing load; at 500 mN the recovery on unloading ranged from 29 to 67 % and the thermal recovery from 35 to 60 %, whereas imprints produced at 25–50 mN recovered almost completely. The depth scatter between nominally identical indents was considerable, with the deepest indent penetrating up to 120 % deeper than the shallowest at 25 mN; in the most heterogeneous deposits it narrowed with increasing load, identifying it as a signature of splat-scale heterogeneity consistent with the broad grain-size distribution revealed by EBSD. The deposits ranked in nearly inverse order in the two ways: the deposit with the largest fraction of NiTi recovered least on unloading but most on heating, while the deposit with the highest recovery on unloading recovered least on heating, so that both regained a comparable 72–79 % of the maximum depth by different routes. The most porous deposit recovered least through both elastic and thermal recoveries.

Keywords: Plasma spraying, NiTi alloy, Nanoindentation; healing

# 1. Introduction

Shape memory alloys, NiTi in particular, are in high demand for engineering applications and are widely available commercially [1]. Commercial NiTi is produced by vacuum melting and casting followed by thermomechanical processing and machining; because the alloy is difficult to machine, this conventional route is costly and limits the size and geometry of the final components [2, 3]. Near-net-shape processing routes are therefore of particular interest, and thermal spraying has emerged as one of them: thick NiTi layers have been produced by radio-frequency plasma spraying [4-7] as well as by related processes such as high-velocity air-fuel spraying [8]. In plasma spraying, the powder particles are injected into the plasma, where temperatures of 8000 - 12 000 K are reached, and are fully or partially melted depending on their trajectory [9]. The deposit is built up layer by layer from splats that spread and solidify rapidly on impact, and its properties are governed by the splat microstructure and by the quality of bonding between new splats and the underlying layer [10]. Rapid solidification, incomplete melting and oxidation make the deposit chemically and microstructurally heterogeneous, which may affect both its mechanical and its shape memory behaviour [4]. However, heterogeneity can be reduced, though not eliminated, by tuning the plasma spraying parameters [11, 12]. In our previous works, thick NiTi deposits produced by radio-frequency plasma spraying were characterised in detail and shown to retain shape memory behaviour [4, 5, 7, 13]. The functional trade-off is microstructural heterogeneity, as each sprayed splat melts and solidifies rapidly, producing a lamellar structure in which martensite, austenite and R-phase coexist with intermetallics such as $Ni_4Ti_3$ and $NiTi_2$ and with oxide phases, distributed non-uniformly and accompanied by porosity [4, 12, 14]. Since the transformation temperatures of NiTi are highly sensitive to local composition and oxygen content – a Ni shift of ~0.1 at.% shifts them by ~10 K [15] - the functional state of plasma-sprayed NiTi can vary from point to point, and a single bulk measurement averages over regions that may behave very differently.

Spatial variability of such extent makes local mechanical characterisation necessary. Bulk techniques such as tensile testing, thermomechanical analysis or macrohardness return a single averaged response and cannot attribute behaviour to individual microstructural constituents. Depth-sensing nanoindentation instead probes a micrometre-scale volume and, through repeated placement, yields the distribution of the local response across a heterogeneous surface [16, 17]. It also gives direct access to the recovery of localised surface deformation, which is the functionally relevant quantity because wear, erosion and impact damage initiate at the surface [18]. An indentation imprint in NiTi recovers by two routes. Upon unloading, part of the penetration is recovered immediately through elastic springback and, in transforming NiTi, through reversion of the stress-induced martensite formed beneath the tip. Upon subsequent heating above the austenite finish temperature $A_f$, the shape memory effect recovers a further part of the residual depth, an effect described as thermally induced self-healing [19]. Both channels have been quantified for dense commercial superelastic and shape memory NiTi sheets [18]. In our earlier work on two plasma-sprayed NiTi deposits [14], both elastic and thermal recoveries were evaluated as well: annealing recovered up to approximately 60 % of the residual imprint depth, the larger recovery being attributed to the deposit with the higher martensite fraction, while the imprints made at the lowest loads could no longer be resolved after annealing. That study, however, compared only two deposits sprayed under similar conditions and did not evaluate the scatter between nominally identical indents, which in a heterogeneous deposit may be as informative as the mean response. Whether the balance between recovery on unloading and on heating differs systematically between deposits produced under different spraying conditions, and how it relates to their phase content, porosity and grain structure, has not been established.

The present work addresses these questions. Four NiTi thick coatings were deposited by radio-frequency plasma spraying at plasma powers of 9 and 12 kW and powder feed rates of 2.1 and 4.2 g/min and separated from their substrates to obtain free-standing deposits. They were indented with a 10 μm conospherical tip at five loads between 25 and 500 mN, and the same imprints were profiled by confocal microscopy before and after annealing above $A_f$. The recovery on unloading and the thermal recovery were evaluated for each individual indent together with their scatter, and the reliability and room-temperature stability of the optical depth measurement were verified independently. To relate the scatter to the microstructure, the grain-size distribution in the vicinity of the indents was determined by electron backscatter diffraction (EBSD) on a selected coating, and the distribution of hardness and reduced modulus was mapped by high-speed nanoindentation. The results are related to the phase composition [4], transformation temperatures, porosity and grain structure of each deposit and compared with the commercial-sheet reference [18] and with our earlier results on plasma-sprayed NiTi [14]. The study thereby extends the characterisation of plasma-sprayed NiTi [4, 5] into the local, load-dependent regime and provides a reference dataset for the surface recovery of powder-route NiTi.

# 2. Materials and Experimental Methods

## 2.1. Material and sample preparation

Four thick coatings of bulk NiTi (samples, 1–4) were produced by thermal (RF-inductively coupled) plasma spraying of NiTi powder (TekSpray-15 RF-ICP system, Tekna, Canada, at the Institute of Plasma Physics of the Czech Academy of Sciences, Prague). The feedstock was gas-atomised near-equiatomic NiTi powder (50 at.% Ni, 99.5 % purity, 20–63 μm, American Elements, Los Angeles, CA, USA). The four samples 1–4 differed in plasma powers considered for spraying, feed rate and number of passes and net powder of spraying and the substrate surface without and with grit blast, summarised in Table 1. The sprayed coatings thickness of 0.3–0.5 mm was sufficient to be separated from the substrate and handled as bulk material; substrate and interface effects are therefore outside the scope of this work.

Table 1. Plasma-sprayed parameters for studied samples 1-4. (Chamber pressure: 15 Pa, stand-off distance: 34 mm, Central gas Ar: 10 sccm, Carrier gas Ar: 8 sccm, pre-heat time: 60 s)

| Samples | Substrate | Plasma power, [kW] | Feed rate [g/min] | Spraying setup plan (time [s]*number of spraying ) | Total powder spraying time [s] |
|---|---|---|---|---|---|
| 1 | Stainless steel (grit blast) | 12 | 4.2 | 30*6 | 180 |
| 2 | Stainless steel | 9 | 2.1 | 25*6 | 150 |
| 3 | Stainless steel (grit blast) | 9 | 2.1 | 25*6 | 150 |
| 4 | Stainless steel | 12 | 2.1 | 25*6 | 150 |

For nanoindentation, one face of each sample was ground and metallographically polished down to 1 μm diamond paste to a finish suitable for depth-sensing indentation. As-sprayed surfaces were too rough for reliable local indentation and were not measured. The preliminary characterization of the samples was partly published in our previous article [4]. However, nanoindentation and subsequent recovery are the new findings here.

## 2.2. Nanoindentation

Nanoindentation was carried out in room-temperature as load-controlled experiment using the NanoTest apparatus (P3, MicroMaterials, Wrexham, UK) fitted with a conospherical Rockwell diamond indenter of nominal tip radius 10 μm. Five maximum loads were applied on each sample: 500, 250, 100, 50 and 25 mN. At each load, ten indents with a spacing of 75 μm were placed in a row, which were also separated by 75 μm, creating a matrix of indents from the deepest on the left to the shallowest on the right in the rectangular area 675 μm x 300 μm. Loading/unloading time, along with holding at maximum load (dwell), was set to 10 seconds. The full loading-unloading cycle was recorded as a load-displacement curve for every indent.

The recovery after unloading of each indent was obtained directly from its load–displacement curve as

$$\eta_{elastic}\ [\%] = \frac{h_{max}-h_r}{h_{max}} \times 100 \qquad 1$$

where $h_{\text{max}}$ is the maximum penetration depth and $h_{\text{r}}$ is the residual depth at complete unloading (consistent with Eq. 1 of [18]) .

## 2.3. Thermal annealing and recovery of residual imprints

After unloading, the residual depth of each indent was measured by laser-scanning confocal microscopy (LEXT OLS5000, Olympus, Japan) over a matrix of 10 indents and 5 loads. To check for room-temperature ageing, the residual imprints were measured twice before annealing ~ 4 months apart (e.g. July and November 2025); the change between the two measurements was negligible, confirming the imprints were stable at room temperature before annealing.

The samples were then annealed by heating to 150 °C at 10 K/min, holding at 150 °C for 10 min, and cooling naturally to ambient temperature in air. This peak temperature lies above the austenite-finish temperature of the material, ensuring that the reverse martensite-to-austenite transformation responsible for shape recovery was completed during the anneal. The residual depth of each imprint was re-measured by confocal microscopy. Because of the non-symmetric imprint shape, each depth value is the average of two cross-sectional measurements.

The thermal recovery of each imprint was calculated as

$$\eta_{thermal}\,[\%] = \frac{h_{before} - h_{after}}{h_{before}} \times 100 \qquad 2$$

where $h_{\mathrm{before}}$ and $h_{\mathrm{after}}$ are the residual indent depths before and after annealing (consistent with Eq. 2 from [18]). Reporting both $\eta_{\mathrm{elastic}}$ and $\eta_{\mathrm{thermal}}$ allows the partition of recovery between unloading (reverse transformation / pseudoelastic) and heating (shape-memory / stress-induced-martensite reversal) to be compared directly with the commercial-sheet reference data of [18].

## 2.4. Surface roughness

Line roughness parameters Ra, Rq and Rz were determined from the confocal data on the free surface between the imprints.. Roughness was measured at the same three stages as the indent depths – initial, repeat (pre-anneal), and post-anneal – so that any change in surface topography on annealing could be tracked alongside the imprint recovery.

## 2.5. Microstructural, compositional and phase characterisation

The cross-section of the coating with substrate was prepared by cutting from the bulk cross-sectionally using an electric discharge machine. The samples were polished with up to 1 μm diamond paste. The microstructure of the plasma-sprayed samples was investigated by scanning electron microscope (SEM, Tescan FERA 3 (Tescan, Brno, Czech Republic)). The cross-section images of the coating were investigated by both modes of observation using secondary electrons and backscattered electrons. Energy-dispersive X-ray spectroscopy (EDS) was carried out using the EDAX system (EDAX, Ametek Inc., Mahwah, NJ, USA) with an Octane Super 60 mm$^2$ detector to determine the chemical composition from the substrate to the coating. The phase analysis of the coating layers was detected at room temperature with an X'Pert PRO θ-θ powder diffractometer (Malvern, UK) using Bragg–Brentano geometry at 40 mA and 35 kV with Kα Co radiation (average wavelength λ = 0.1790 nm), a focus-slit distance of 100 mm, and a goniometer radius of 240 mm. The data were measured in the 2θ range of 20–120°, using a step size of 0.013°, a scan step time of 1.4 s, and a fixed divergence slit size of 0.5°. The phase transformation temperatures of powder and coating were carried out by differential scanning calorimetry (DSC 25, TA Instruments, New Castle, DE, USA) at a heating and cooling rate of 5 °C/min in the temperature range of −100 to +150 °C in a nitrogen environment inside the sample chamber. The porosity within the cross-section of plasma-sprayed samples was estimated using a Keyence microscope. The phase and grain orientation distribution analyses were performed using the electron back-scattered diffraction method (EBSD) with the EDAX DigiView V camera and system (EDAX, Ametek Inc., Mahwah, NJ, USA). A voltage of 20 kV, a current of 1–2 nA, and a working distance of 15 mm were chosen for analysis. The EBSD sample followed the procedure, as samples were hot-mounted in conductive Bakelite and were metallographically prepared by the modified procedure for nickel (ASTM C-56 recipe). The colloidal silica was used in the last polishing step, and the surface was finished by etching in Kroll's reagent. Grain orientation and phase mapping of the plasma sprayed NiTi alloys were performed by the electron backscatter method.

## 2.6. Nanoindentation mapping of hardness and reduced modulus of NiTi coating by Berkovich indenter

The polished plasma-sprayed samples were used for high-speed nanoindentation (HSNI) at room temperature in Continuous stiffness measurement mode (CSM) using a Berkovich tip. Continuous measurement of contact stiffness is done during loading, which allows measurement of hardness and elastic modulus as a function of indentation depth. HSNI is used to map mechanical properties such as hardness and elastic modulus at spatial resolution and high speed across the surface of materials (especially at micro- or nanoscale). HSNI indentations were performed in a grid pattern across the sample's surface.

The evaluation processes of indentation modulus and hardness are derived from Oliver and Pharr theory [20]. As per the Oliver-Pharr equation, reduced modulus is given by Eq.3.

$$E_r = \frac{\sqrt{\pi}}{2\beta}\frac{S}{\sqrt{A_p}} \quad 3$$

here $E_r$ is reduced modulus, $\beta$ is tip shape correction factor (1.034 for Berkovich tip), $A_p$ is projected contact area and $S$ is dynamically measured contact stiffness in CSM mode.

## 2.7. Structural evidence by sub microstructural investigation

Samples for TEM observation of lattice defects in the thermal plasma sprayed samples 1, 2 were prepared by the Focused Ion Beam (FIB) method using the FEI Quanta 3D FIB-SEM electron microscope. TEM observations were carried out using the FEI Tecnai TF20 X-twin field emission gun microscope operated at 80 kV. EDX spectroscopy, TEM, STEM and electron diffraction were used for imaging and analysis of the TEM lamellae.

# 3. Results and Discussion

The microstructure, porosity and phase composition of the deposits are described first, followed by the elastic and the thermal recovery of the indents. Certain inhomogeneities were observed during the measurements, which will be discussed with the aid of microscopic images in the chapter on thermal recovery.

## 3.1. Characterization of coatings: Microstructural examination

Fig. 1 displays the cross section of coating layers with the stainless steel. The SEM images were presented in SE and BSE mode to observe the interfacial adhesion of coating layers with stainless steel and also to observe the coating layers along the cross section.

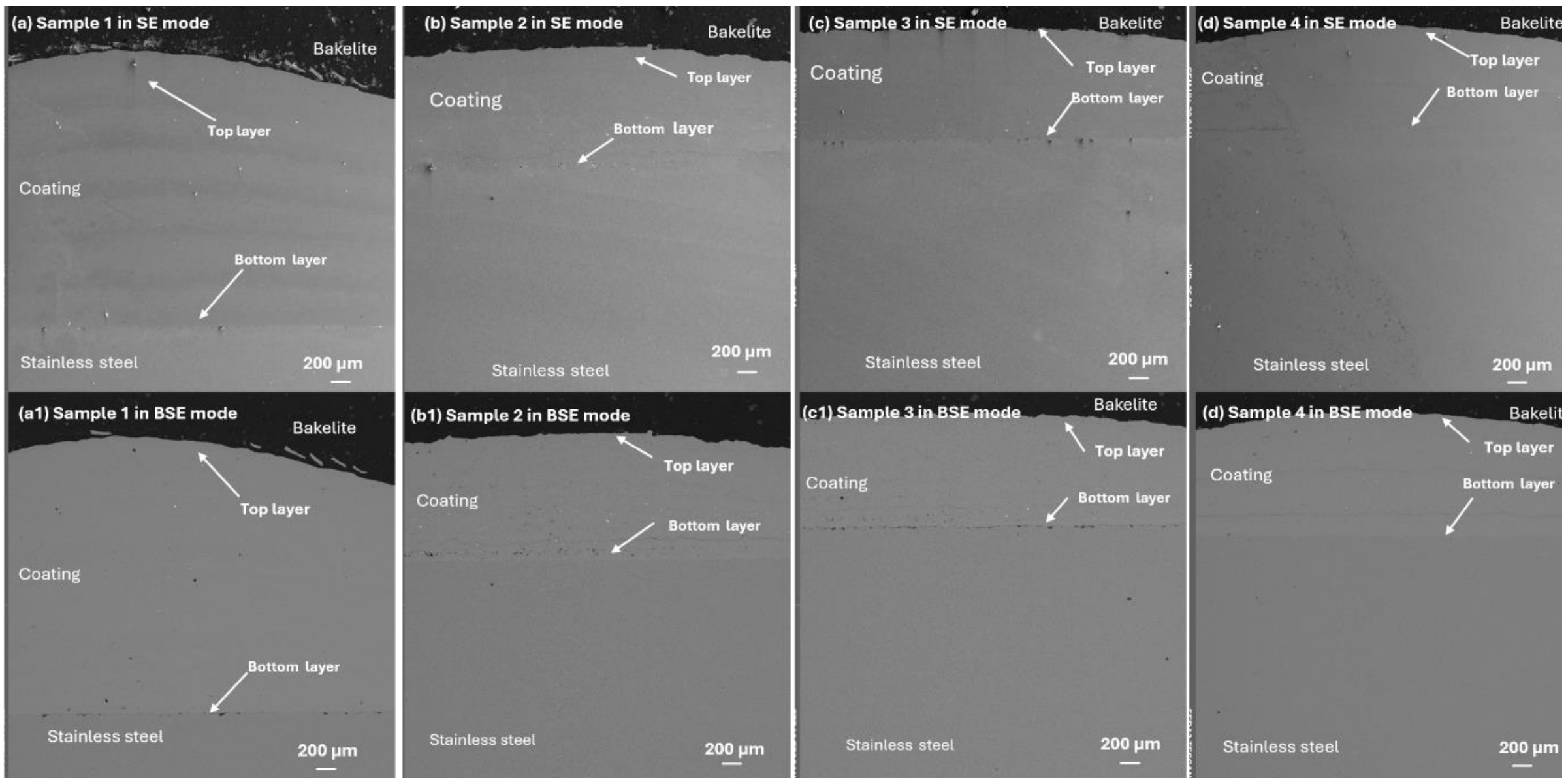

Fig. 1. Cross-section images of coating on substrate present in both SE and BSE mode with indication of top and bottom layer.

The porosity of the deposits, determined by image analysis of polished cross-sections, is summarised in Table 2. Porosity is located mainly at the coating–substrate interface in samples 1 and 2, and between or within the spray layers. In overall, samples 1–3 show rather similar porosity of 12.6–13.6 %; given standard deviations of 3–4 %, the differences between them are not significant. Sample 4 is markedly more porous (18.7 ± 1.5 %). The porosity does not follow the spraying parameters in a simple way: Sample 4 was sprayed at the higher plasma power of 12 kW, which would generally be expected to promote more complete melting and a denser deposit. Pores reduce the load-bearing volume beneath the indenter and can collapse irreversibly under load; their contribution to the residual depth is therefore expected to lower both the elastic and the thermal recovery, as discussed in Chapters 3.3 and 3.4.

Table 2. Total area of deposition, total area of porosity and Porosity (%) in plasma-sprayed samples.

| Sample | Total Deposit Area [mm²] | Total Pore Area [mm²] | Porosity [% ± Std] |
|---|---|---|---|
| Sample 1 | 117.85 | 15.98 | 13.56% ± 3.2% |
| Sample 2 | 38.55 | 4.85 | 12.57% ± 4.1% |
| Sample 3 | 31.93 | 4.20 | 13.15% ± 3.2% |
| Sample 4 | 21.78 | 4.07 | 18.66% ± 1.5% |

## 3.2. Phase analysis and Transformation of Phases in NiTi powder and plasma sprayed samples

The transformation behaviour of NiTi powder and the four deposited coatings 1–4 during the heating and cooling range is shown in Fig.2. The transformation temperature shows the austenite finish temperature of NiTi particles shifts towards higher temperature from 62.5 ° towards plasma sprayed coating up to 105.2 °C. The samples are completely in the austenite phase above $A_f$ finish temperature.

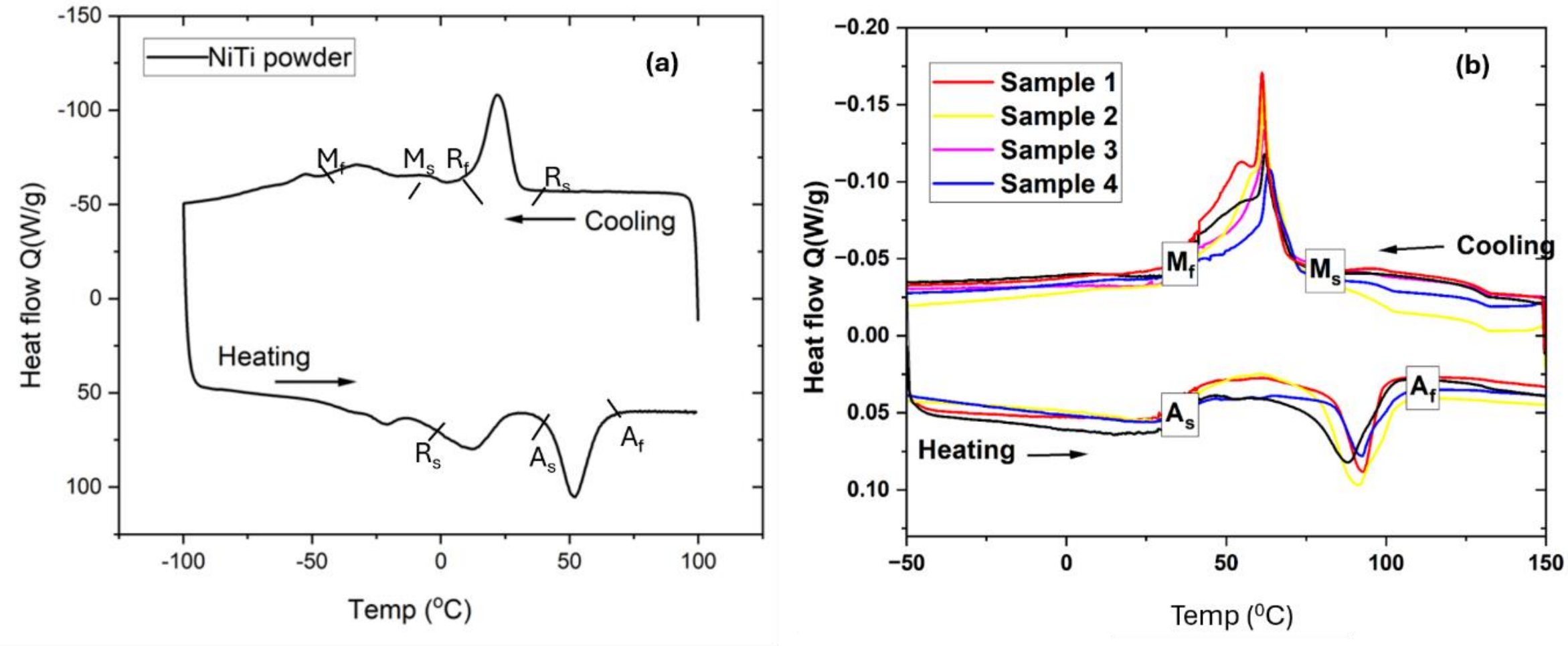

Fig. 2. Transformation temperatures of austenite and martensite phase (start, finish) for NiTi powder and coatings 1-4.

The upward shift is consistent with a depletion of Ni in the NiTi matrix during spraying [8]. The DSC curves of the four deposits nearly coincide, so that the transforming NiTi matrix is martensitic at room temperature in all of them, and the annealing temperature of 150 °C lies well above $A_f$ for every sample. The phase analysis revealed the presence of austenite B2, martensite B19', intermetallic compound $NiTi_2$ and $Ni_4Ti_3$, oxide phases such as $Ti_4Ni_2O$ and metallic Ti. Fig. 3 displays the XRD peak profile as a function of angle position in the various samples. Table 3 displays the quantitative evaluation of various phases in the plasma-sprayed samples.

Table 3. Quantitative evaluation of various phases in coatings 1–4.

| Phases | NiTi powder | Sample 1 | Sample 2 | Sample 3 | Sample 4 |
|---|---|---|---|---|---|
| SemiQuant % (w/w) | | | | | |
| Austenite NiTi (B2) | 64 | 48.0 | 43.2 | 27 | 39.5 |
| Martensite NiTi (B19') | 36 | 39.9 | 25.8 | 39 | 26.0 |
| Oxide phase $Ti_4$ $Ni_2$ O | | 3.2 | - | | 2.0 |
| Precipitates $Ni_4Ti_3$ | | - | 30.6 | | - |
| Intermetallic $Ti_2Ni$ | | 6.5 | - | 34 | 32.5 |
| Titanium Ti | | 2.3 | 0.5 | - | - |

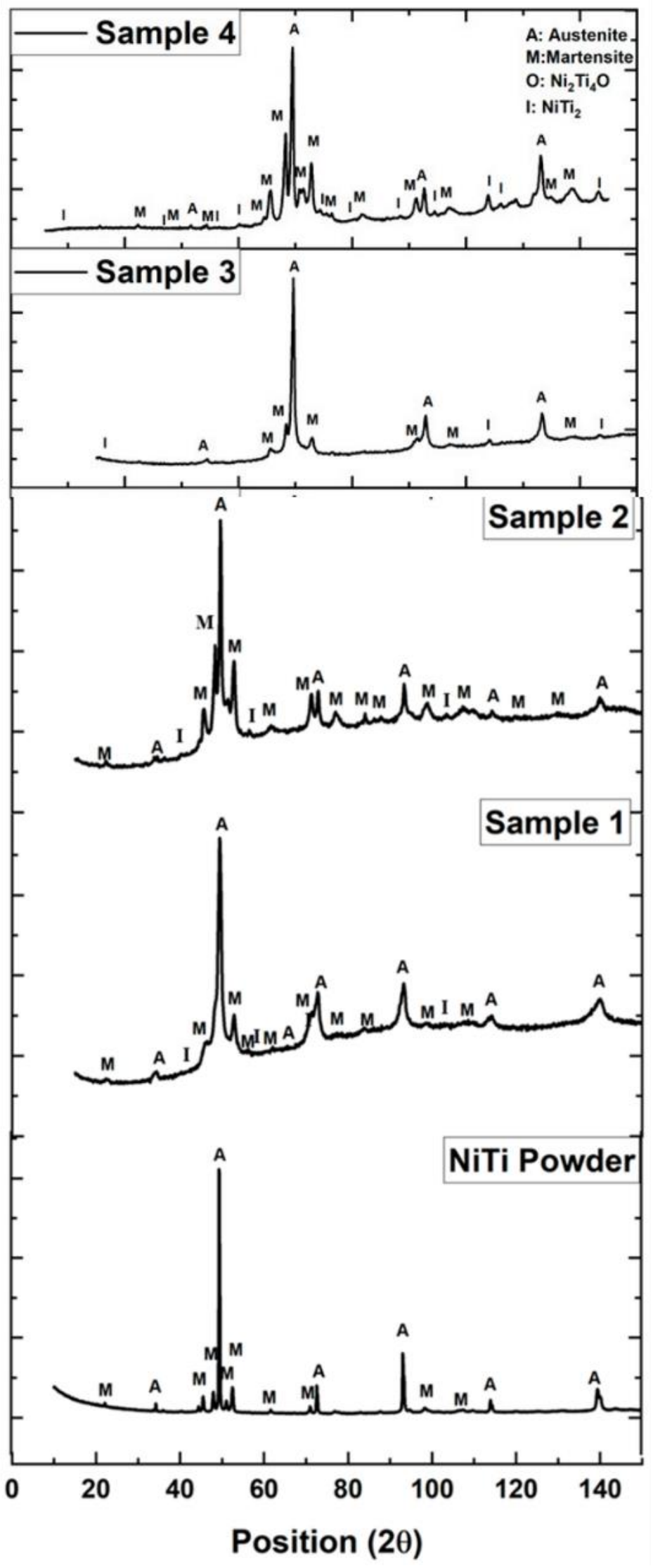

Fig. 3 XRD pattern of NiTi powder and thermal plasma sprayed samples 1-4.

## 3.3. Deformed surface recovery after unloading

Indentation under various loads of 25, 50, 100, 250 and 500 mN led to indents of various depths that were mostly homogeneous, except for several indents. Fig. 4 – Fig. 7 show the load-displacement curves for all four samples with various loads, with all ten indents per sample overlaid.

Whereas dense commercial NiTi sheets produce near-coincident curves under repeated indentation at a given load [18], the present plasma-sprayed samples show a pronounced spread in penetration depth at identical load. The effect is largest at low loads: in sample 1 at 25 mN the deepest indent penetrated 427 nm against 197 nm for the shallowest, nearly 120 % deeper, corresponding to a standard deviation of 28 % of the mean, with sample 3 and sample 2 showing comparable though smaller spreads of 17 % and 13 % (Tab. 4). This spread is the signature of the lamellar, multi-phase microstructure produced by the spray process, in which individual indents sample different local combinations of martensite, austenite, intermetallic and oxide phases together with porosity [4, 14].

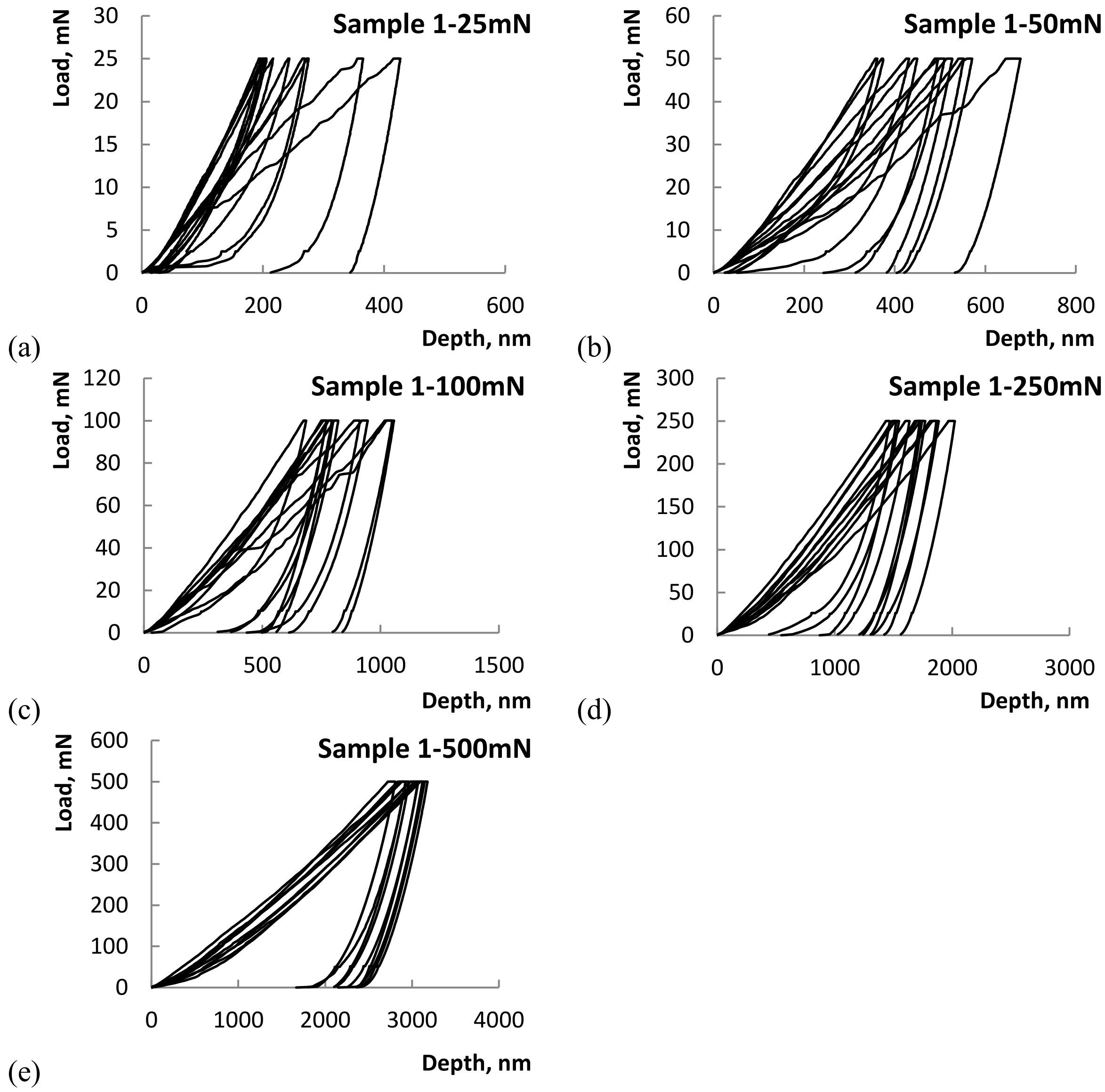

Fig. 4. Load-displacement curves of 10 overlaid indents with loads of (a) 25 mN, (b) 50 mN, (c) 100 mN, (d) 250 mN, (e) 500 mN on plasma-sprayed NiTi sample 1.

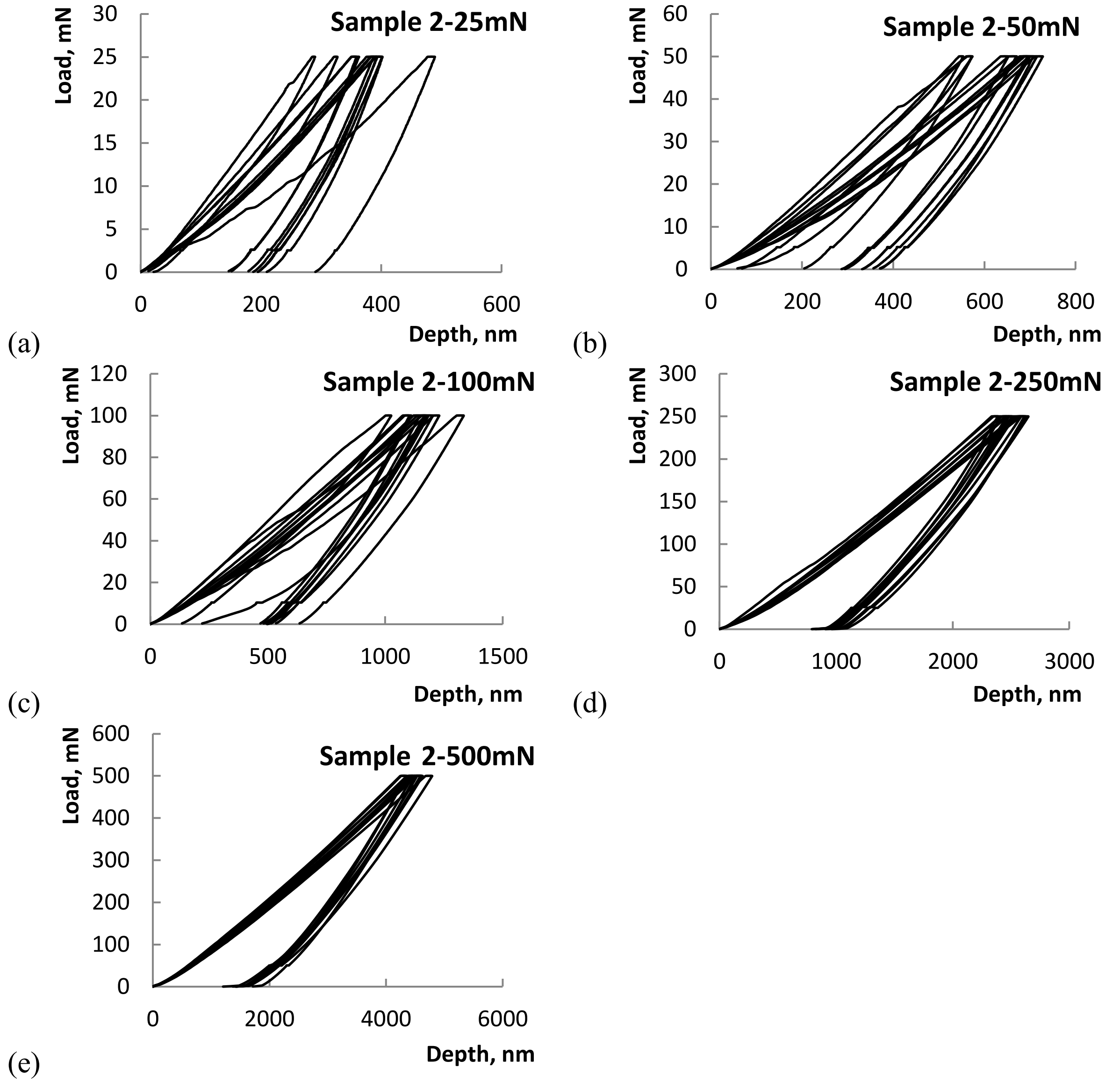

Fig. 5. Load-displacement curves of 10 overlaid indents with loads of (a) 25 mN, (b) 50 mN, (c) 100 mN, (d) 250 mN, (e) 500 mN on plasma-sprayed NiTi sample 2.

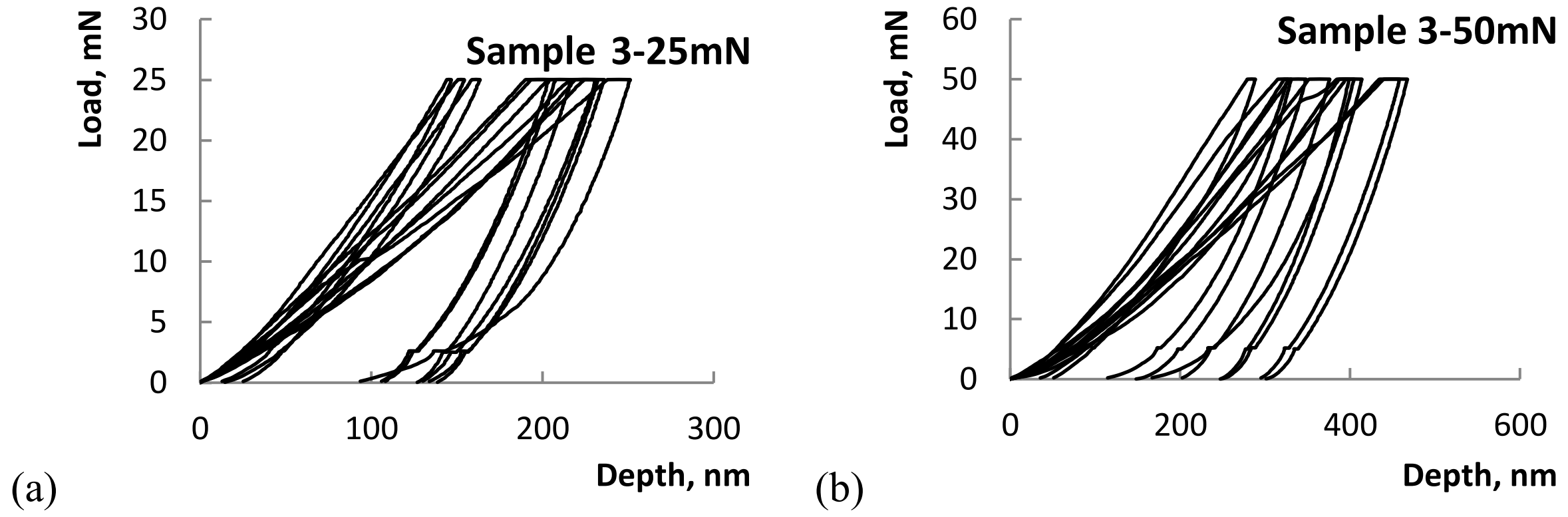

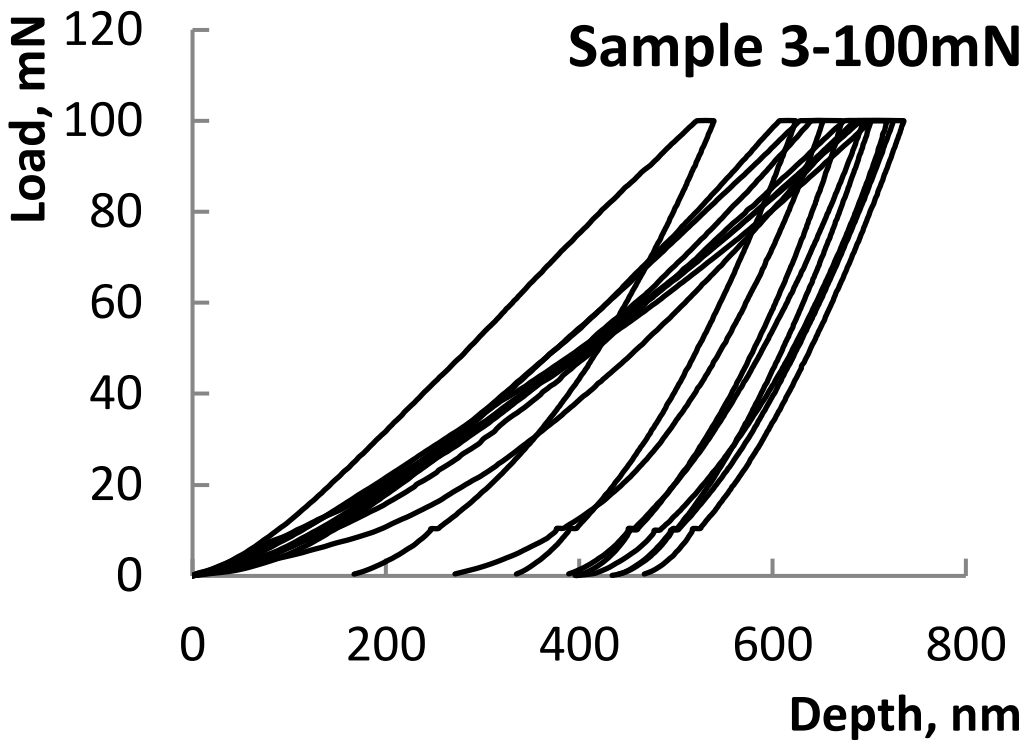

(c)

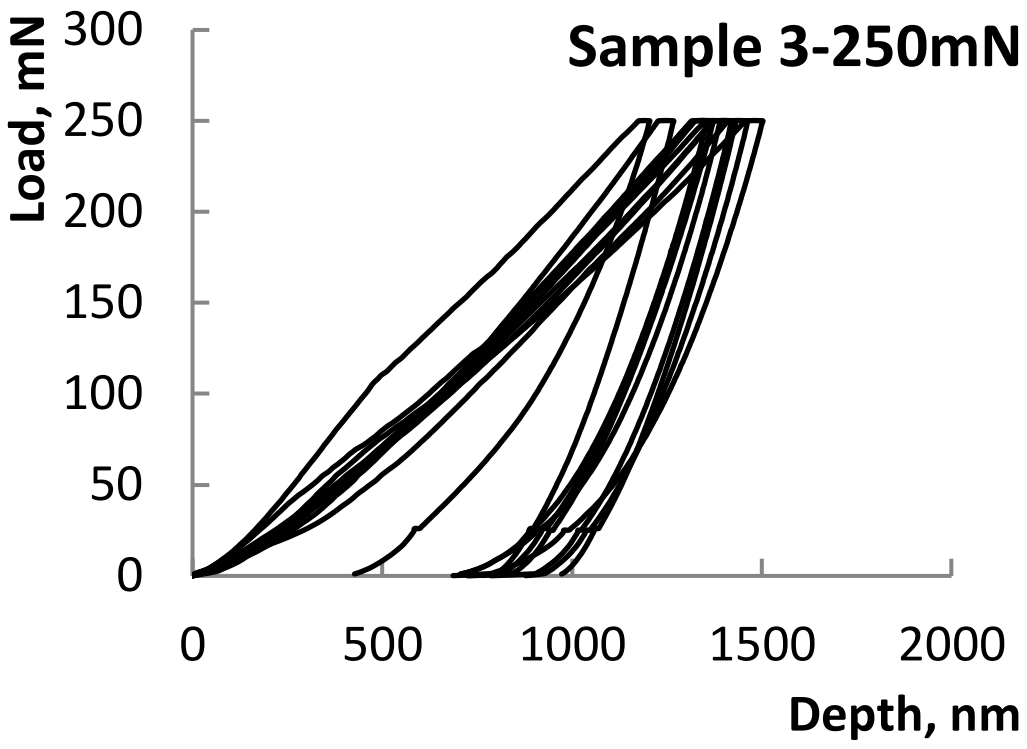

(d)

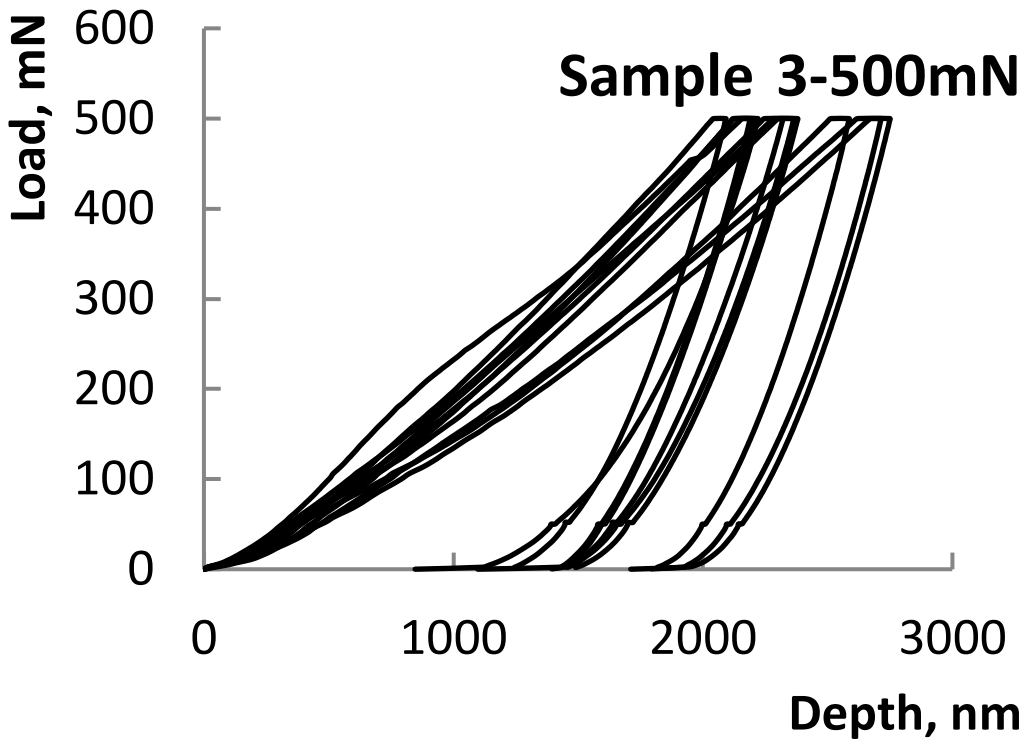

(e)

Fig. 6. Load-displacement curves of 10 overlaid indents with loads of (a) 25 mN, (b) 50 mN, (c) 100 mN, (d) 250 mN, (e) 500 mN on plasma-sprayed NiTi sample 3.

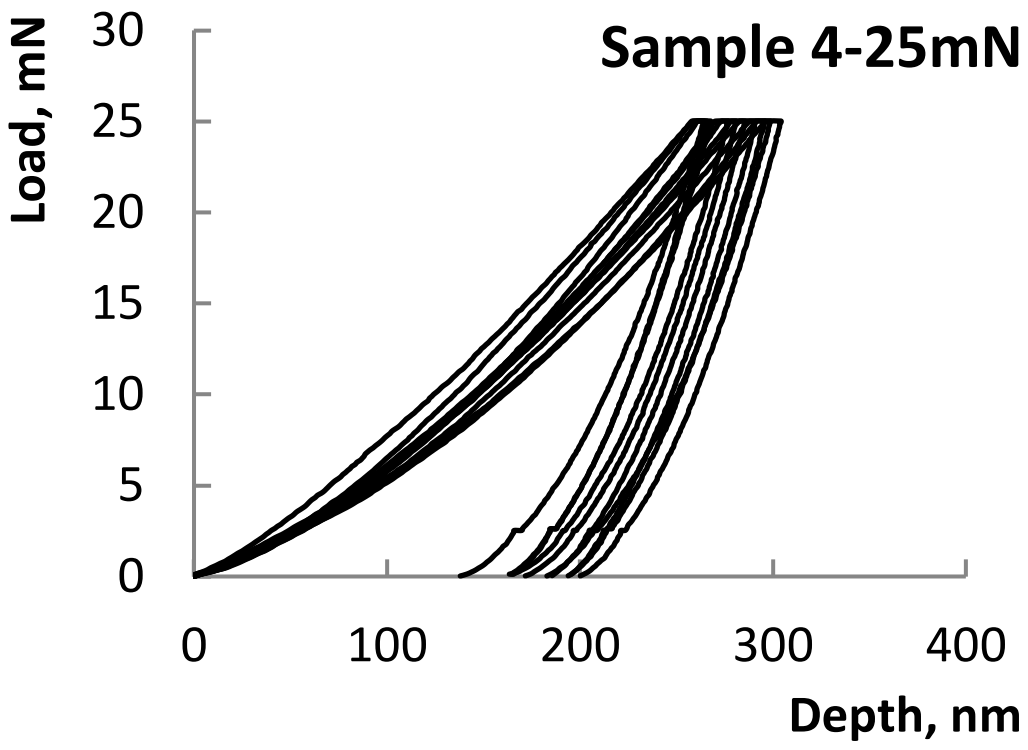

(a)

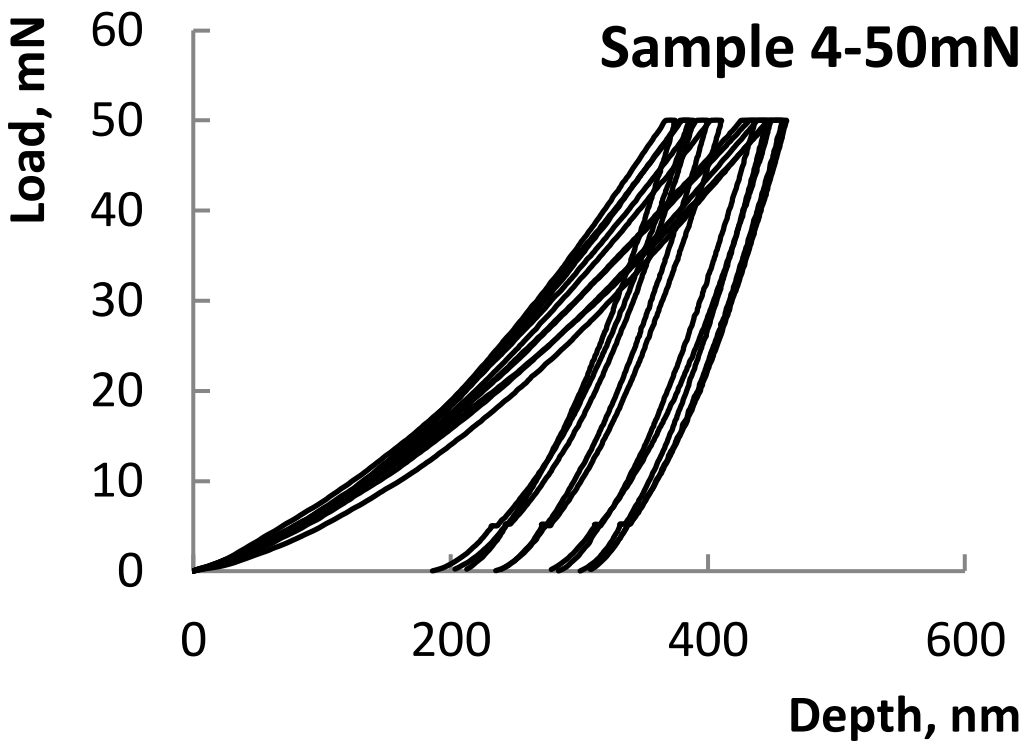

(b)

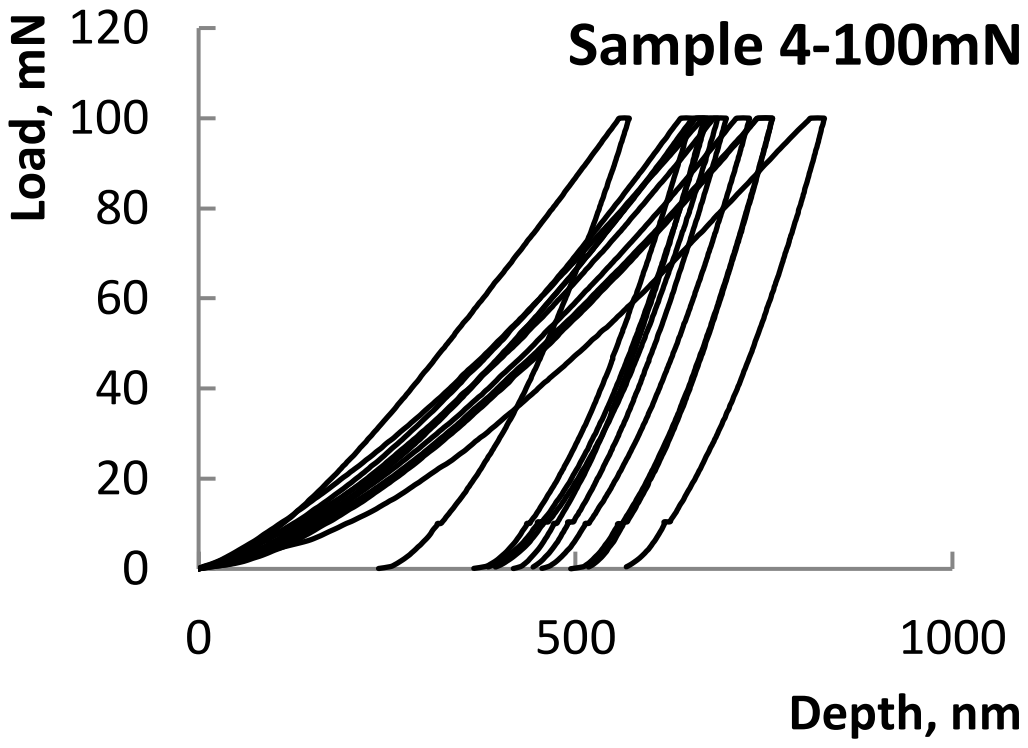

(c)

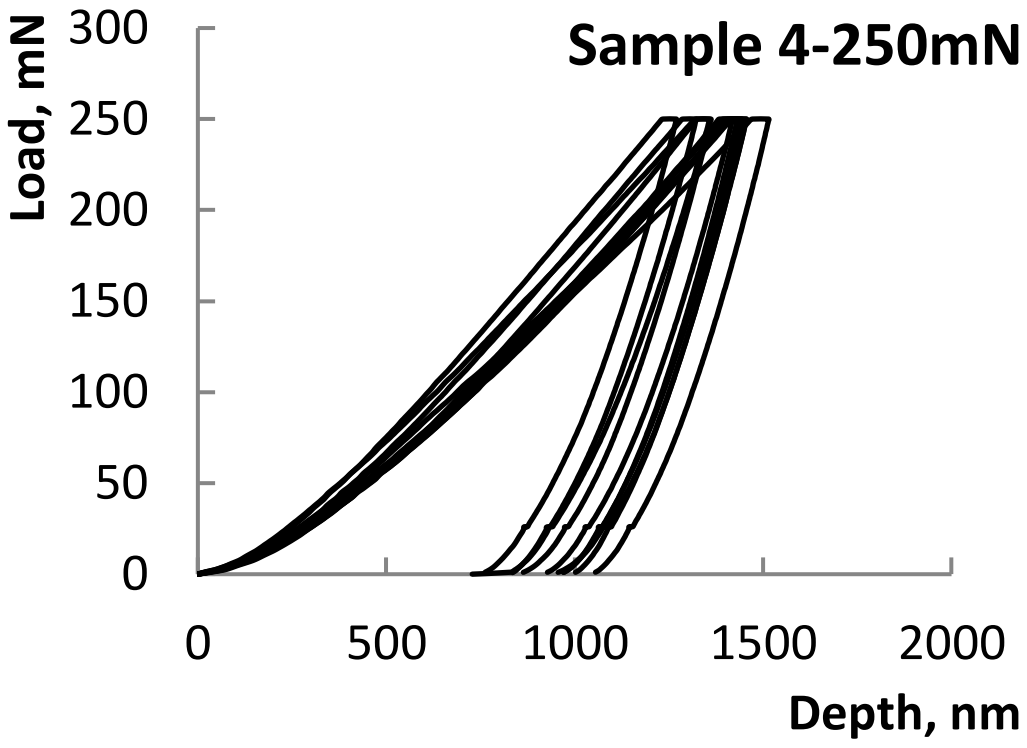

(d)

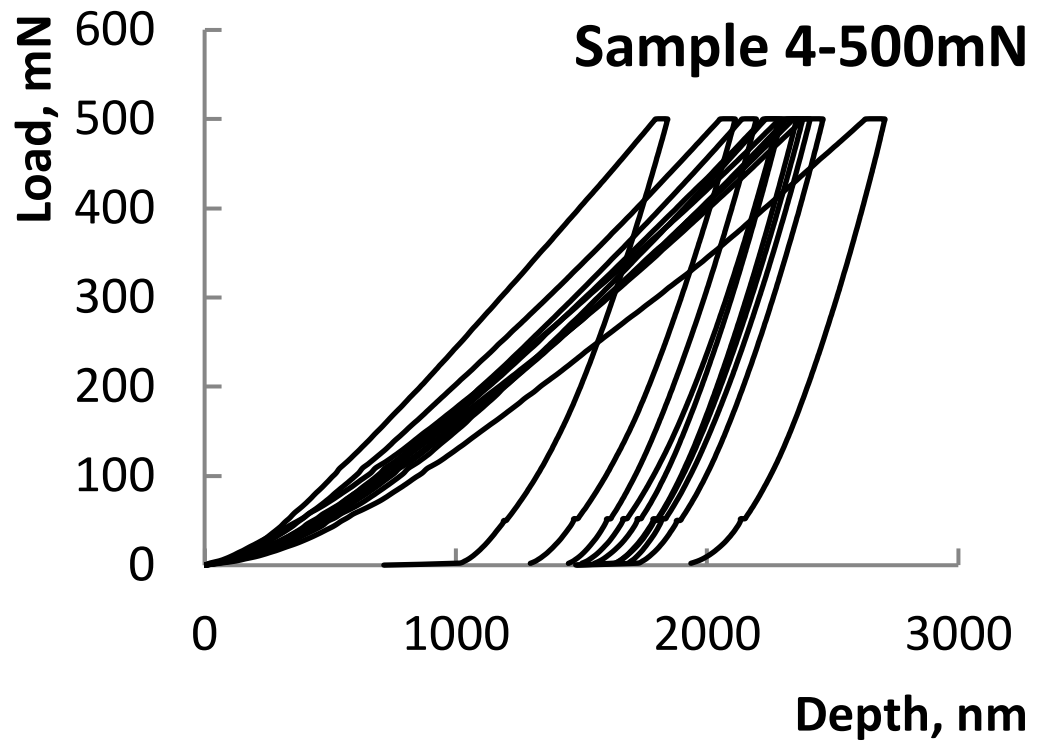

(e)

Fig. 7. Load-displacement curves of 10 overlaid indents with loads of (a) 25 mN, (b) 50 mN, (c) 100 mN, (d) 250 mN, (e) 500 mN on plasma-sprayed NiTi sample 4.

Table 4. Averages of maximum loads $h_{max}$ of 10 indents from all measured loads of samples 1-4

| | Sample 1 [nm] | Sample 2 [nm] | Sample 3 [nm] | Sample 4 [nm] |
|---|---|---|---|---|
| 25 mN | 261 ± 74 | 380 ± 50 | 204 ± 35 | 283 ± 14 |
| 50 mN | 495 ± 91 | 657 ± 63 | 381 ± 55 | 421 ± 31 |
| 100 mN | 861 ± 118 | 1168 ± 78 | 810 ± 415 | 704 ± 67 |
| 250 mN | 1719 ± 168 | 2525 ± 87 | 1377 ± 84 | 1403 ± 70 |
| 500 mN | 3034 ± 115 | 4540 ± 127 | 2382 ± 216 | 2304 ± 217 |

Table 5. Averages of the recovery after unloading $\eta_{elastic}$ from all measured loads of samples 1-4.

| | Sample 1 [%] | Sample 2 [%] | Sample 3 [%] | Sample 4 [%] |
|---|---|---|---|---|
| 25 mN | 79 ± 25 | 60 ± 18 | 59 ± 20 | 38 ± 4 |
| 50 mN | 56 ± 30 | 60 ± 15 | 55 ± 18 | 40 ± 6 |
| 100 mN | 45 ± 21 | 62 ± 11 | 43 ± 14 | 40 ± 7 |
| 250 mN | 38 ± 16 | 64 ± 3 | 46 ± 9 | 35 ± 5 |
| 500 mN | 29 ± 6 | 67 ± 4 | 40 ± 9 | 37 ± 9 |

The elastic recovery ($\eta_{elastic}$) (see Tab.5.) decreases with increasing load in two of the four samples: most markedly in Sample 1, from 79 ± 25 % at 25 mN to 29 ± 6 % at 500 mN, and more moderately in Sample 3, from 59 ± 20 % to 40 ± 9 %, while Sample 4 remains approximately constant at 35 – 40 % across the whole load range. Also, Sample 2 has rather similar values of 60 – 67 % in the same range. This load dependence is expected. At low loads, the deformed volume is accommodated predominantly by elastic deflection and by stress-induced martensitic transformation beneath the tip, both of which reverse upon unloading. As the load increases, these contributions saturate while the irreversible contributions – dislocation plasticity and the collapse of porosity – continue to grow, so that the recoverable fraction of the total penetration is less prominent.

The scatter in $\eta_{elastic}$ follows the same pattern as that in $h_{max}$ and is itself informative (see Tab. 4 and Tab. 5). Sample 1 shows by far the largest uncertainty at low loads (± 25 and ± 30 % at 25 and 50 mN), with individual indents ranging from essentially complete recovery to values comparable with the other samples; its mean values in this range therefore do not permit a meaningful comparison between samples. The scatter narrows progressively with load, reaching ± 6 % at 500 mN. The ranking by heterogeneity is itself load-dependent: whereas sample 1 is the most heterogeneous, the other three samples 2–4 show markedly more uniform behaviour at all loads, with the most uniform sample 4 (± 4 to ± 9 %). Load dependence of scatter can be explained by the fact that at the lowest loads the indented volume is comparable in size with the individual splats and phase domains formed during spraying, so that each indent probes an essentially distinct local microstructure; as the load increases, the deformed volume averages over a growing number of such domains and the response converges. This load dependence of the scatter, together with the metallographically polished surface finish, confirms that the spread originates in the material rather than in the measurement and provides a quantitative measure of the microstructural heterogeneity of each deposit.

Sample 2 differs from the other three samples in two respects. It exhibits the highest elastic recovery at every load, and it is the only sample in which the recovery does not decrease with increasing load, rising slightly from 60 ± 18 % at 25 mN to 67 ± 4 % at 500 mN. It is also the most compliant sample, reaching the largest penetration at every load (with 10 indents averaging 4540 ± 127 nm at 500 mN, against 2304 ± 217 nm for Sample 4). The combination of high compliance with high recoverability is not predicted by hardness alone, since a more compliant material would ordinarily accumulate a larger unrecovered depth; it is, however, consistent with a reversible, transformation-related deformation beneath the tip rather than with dislocation glide. The opposite behaviour is shown by Sample 4, which is among the least compliant samples and shows the lowest elastic recovery at every load, indicating a deformed volume in which a substantial fraction of the material does not participate in the reversible transformation.

A single pronounced exception occurs for sample 3 at 100 mN, where $h_{max}$ = 2043 nm, while other indents range from 539 to 735 nm with rather low spread. Higher loads of 250 mN and 500 mN also have very low spread of data, 1204–1502 nm and 2092–2749 nm, respectively. Such isolated deep penetration is attributed to an indent ± 25 % most probably placed above a subsurface pore or incompletely bonded splat boundary, where the material beneath the tip collapses locally. This point is retained in the average and identified, since it can represent a real feature of the deposit rather than a measurement artefact.

## 3.4. Recovery after annealing

Indentation depths were compared prior to and after thermal annealing and measured using a confocal 3D microscope. In the first approach, thermal recovery can be observed visually by comparing 2D intensity microscope images of the impression matrices (see Fig.8). Here, the inhomogeneity of Sample 1, already mentioned in Section 3.1, is apparent. Different shades of surrounding areas in the images are an unfortunate intensity artefact, but do not affect the measurement.

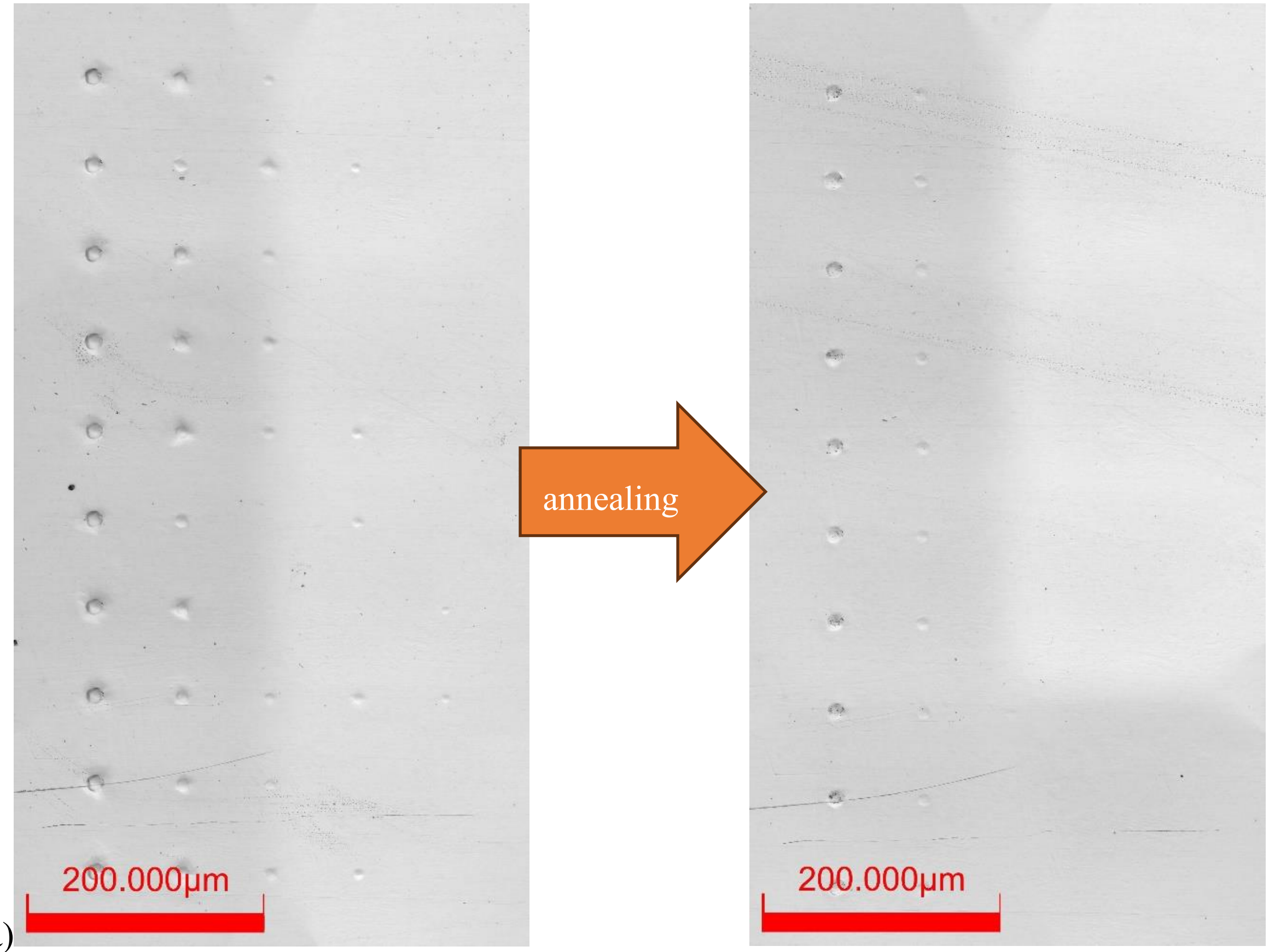

a)

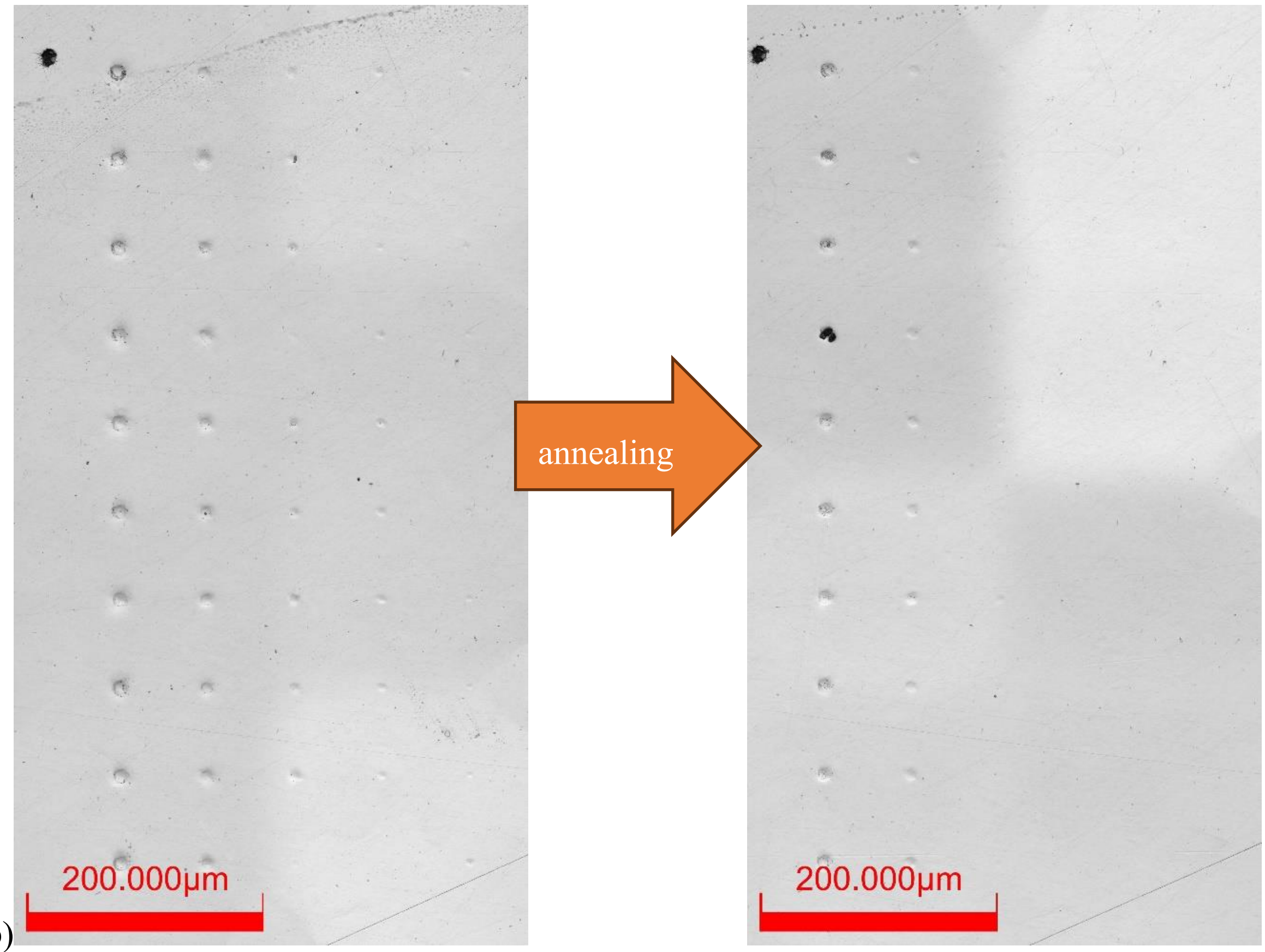
200.000μm
annealing
200.000μm

b)

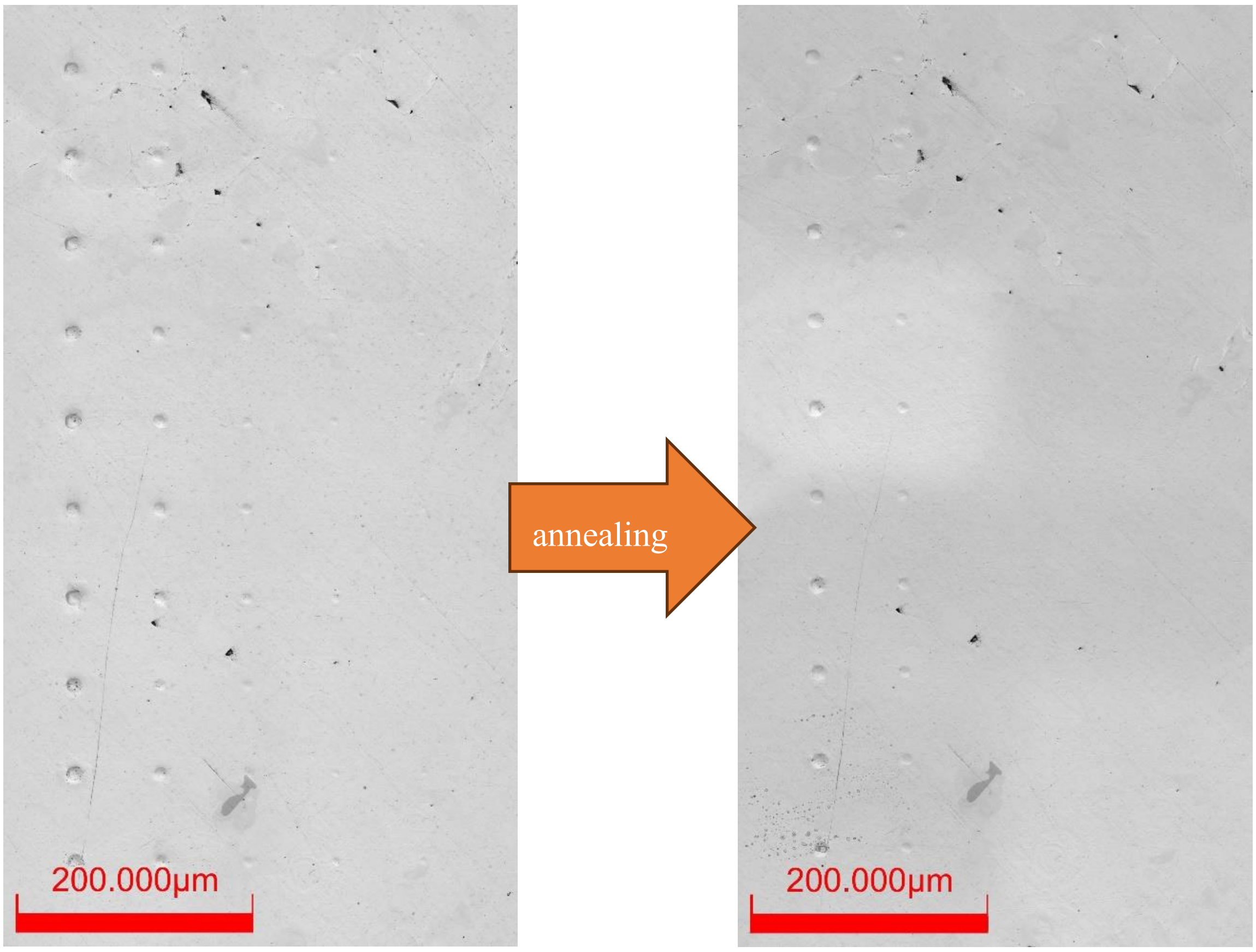
200.000μm
annealing
200.000μm

c)

d)

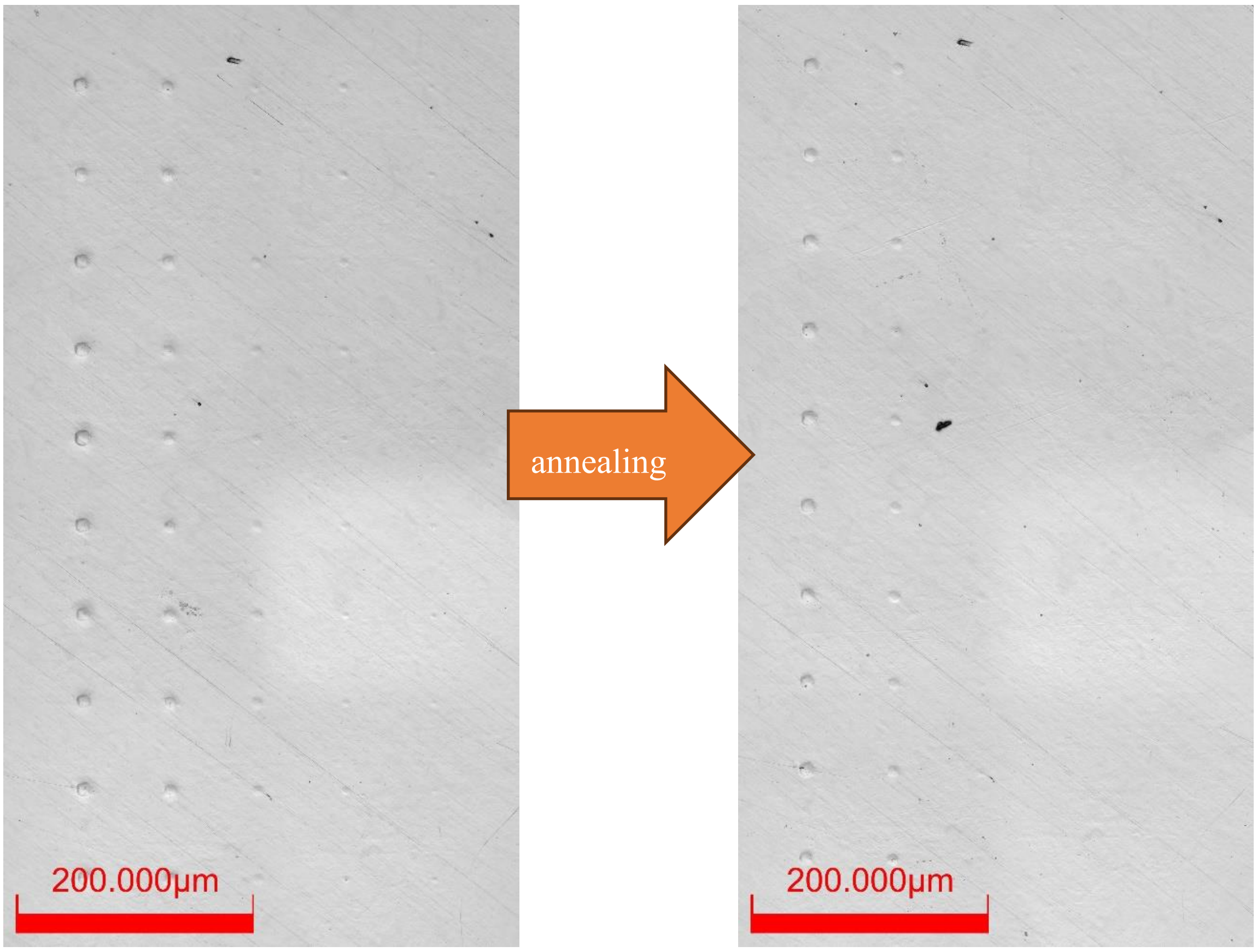

Fig. 8. Confocal microscope intensity images of whole indentation matrices with five loads 10 indents each, for samples a) sample 1, b) sample 2, c) sample 3 and d) sample 4 before and after annealing.

The reliability of the confocal depth measurement was verified by comparing the residual imprint depths measured optically with the residual depths $h_{res}$ obtained independently from the corresponding load-displacement curves on sample 4. The two methods agree closely: the mean residual depths differ by 2–6 % at all five loads, and the per-indent correlation is strong (r = 0.91–0.95 between 50 and 500 mN), with a typical deviation of 24–34 nm. At 25 mN, the mean depths for sample 4 are still in good agreement. However, residual depths $h_{res}$ of sample 1 were so low that microscopic measurement was impossible.

Each imprint was measured twice prior to annealing, separated by an interval of ~ 4 months, in order to verify that the residual deformation did not relax spontaneously at room temperature. The two measurements agree to within a median of 1.5 % for all imprints deeper than 0.1 µm, confirming that the residual imprints are stable and that the recovery reported here is induced by the thermal treatment alone. Thermal recovery was computed by equation (2) using confocal microscope measurements and is summarized in Table 6. While thermally recovered indents for higher loads 500 – 100 mN were easy to find, the 25 mN indents were practically impossible to measure after annealing using confocal microscope, thus, the lines in Tab. 6 report ca. 100 % recovery (meaning recovery is complete or below the detection limit of the confocal microscope). Thermal recovery decreases systematically with increasing indentation load in all four samples. At 500 mN the recovery ranges from 60 ± 2 % in sample 1 down to 35 ± 4 % in sample 2, whereas at 100 mN all four samples recover between 74 and 90 % of their residual depth. The ranking of the samples by thermal recovery is close to the inverse of their ranking by elastic recovery. Sample 1, with the lowest elastic recovery at 500 mN (29 ± 6 %), shows the highest thermal recovery (60 ± 2 %), whereas sample 2, with the highest elastic recovery (67 ± 4 %), recovers the least upon annealing (35 ± 4 %). Combining both channels, the total recovered fraction of the maximum penetration depth is comparable in these two samples, approximately 72 % for sample 1 and 79 % for sample 2, although it is reached by different routes.

Indentation depths are also summarized in Fig.9 before and after annealing.

Table 6. Averages of the thermal recovery $\eta_{thermal}$ from all measured loads of samples 1-4.

| Load / Recovery (%) | Sample 1 | Sample 2 | Sample 3 | Sample 4 |
|---|---|---|---|---|
| 25 mN | immeasurable | < 100 | ~ 100 | ~ 100 |
| 50 mN | ~ 100 | 85 ± 8 | 88 ± 7 | 81 ± 8 |
| 100 mN | 90 ± 8 | 77 ± 10 | 85 ± 8 | 74 ± 10 |
| 250 mN | 74 ± 11 | 58 ± 8 | 61 ± 15 | 52 ± 6 |
| 500 mN | 60 ± 2 | 35 ± 4 | 47 ± 9 | 36 ± 3 |

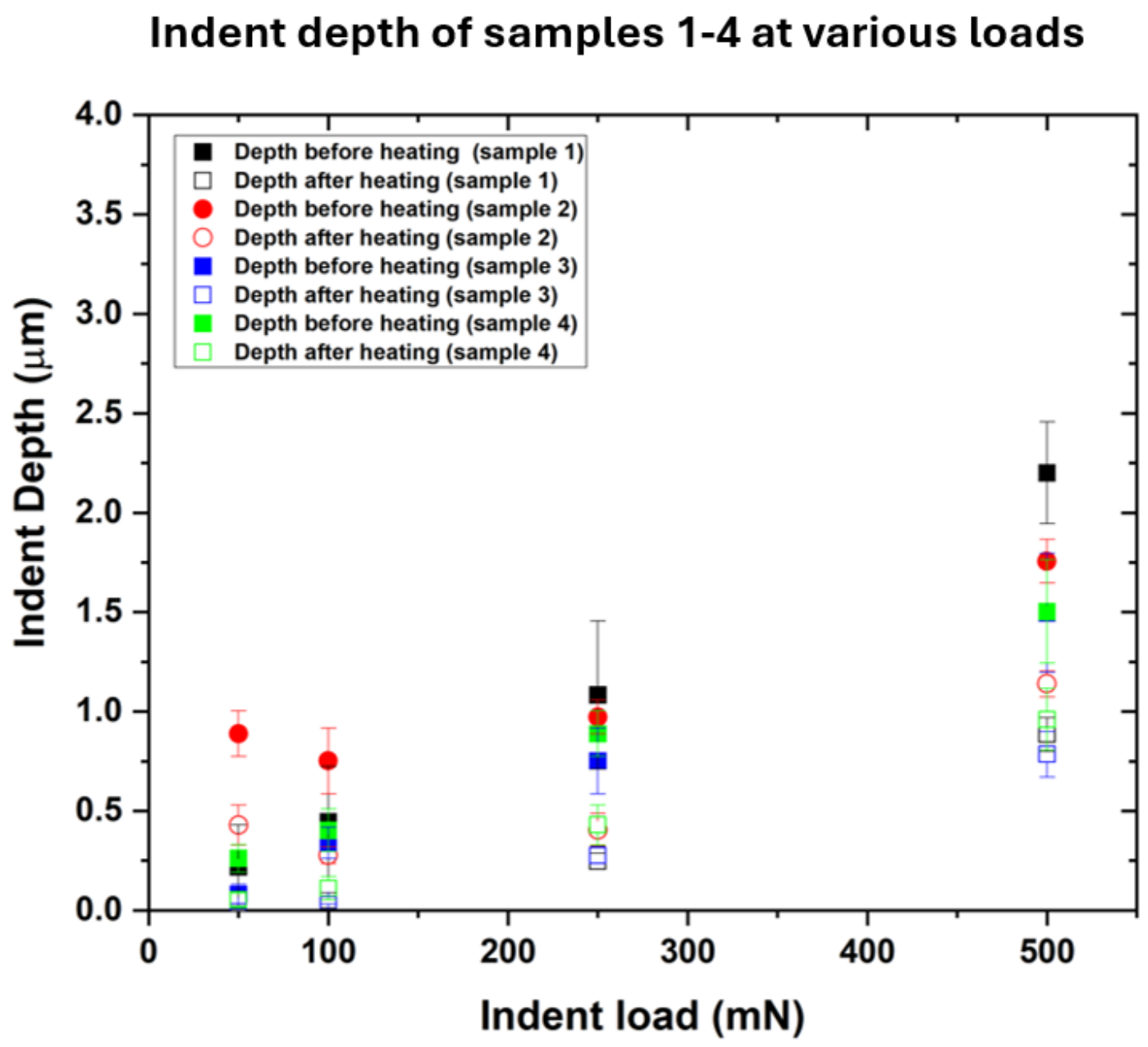

Fig. 9. Indent depth versus indent load before and after heating for samples 1–4.

## 3.5. Roughness measurement

The surface roughness was measured on the free surface between the imprints, so that the values are unaffected by the indentation marks themselves. Two measurements were performed before annealing and were identical within the scatter for all four samples, providing further confirmation that the surface state did not change spontaneously between the two measurement dates (as reported ~ 4 months apart).

Annealing did not change the surface roughness of the deposits significantly (see Fig. 10) as the Ra, Rq and Rz for all four samples lie after annealing within the scatter of the two pre-annealing measurements, the largest decrease being observed for Sample 2 (Ra from 0.024 ± 0.006 to 0.016 ± 0.004 µm). This agrees with the line roughness reported for commercial superelastic and shape memory NiTi sheets [18], where Ra and Rz also changed within the measurement scatter upon annealing; the increase noted there concerned the areal parameter Sz and the waviness-filtered roughness, which were not evaluated here. In our earlier study of plasma-sprayed NiTi [14], the disappearance of the low-load imprints after annealing was ascribed to an increase in surface roughness. The present results, in which imprints that were reliably measurable before annealing disappeared while the roughness remained unchanged, show that their disappearance reflects essentially complete recovery rather than masking by surface roughness.

.

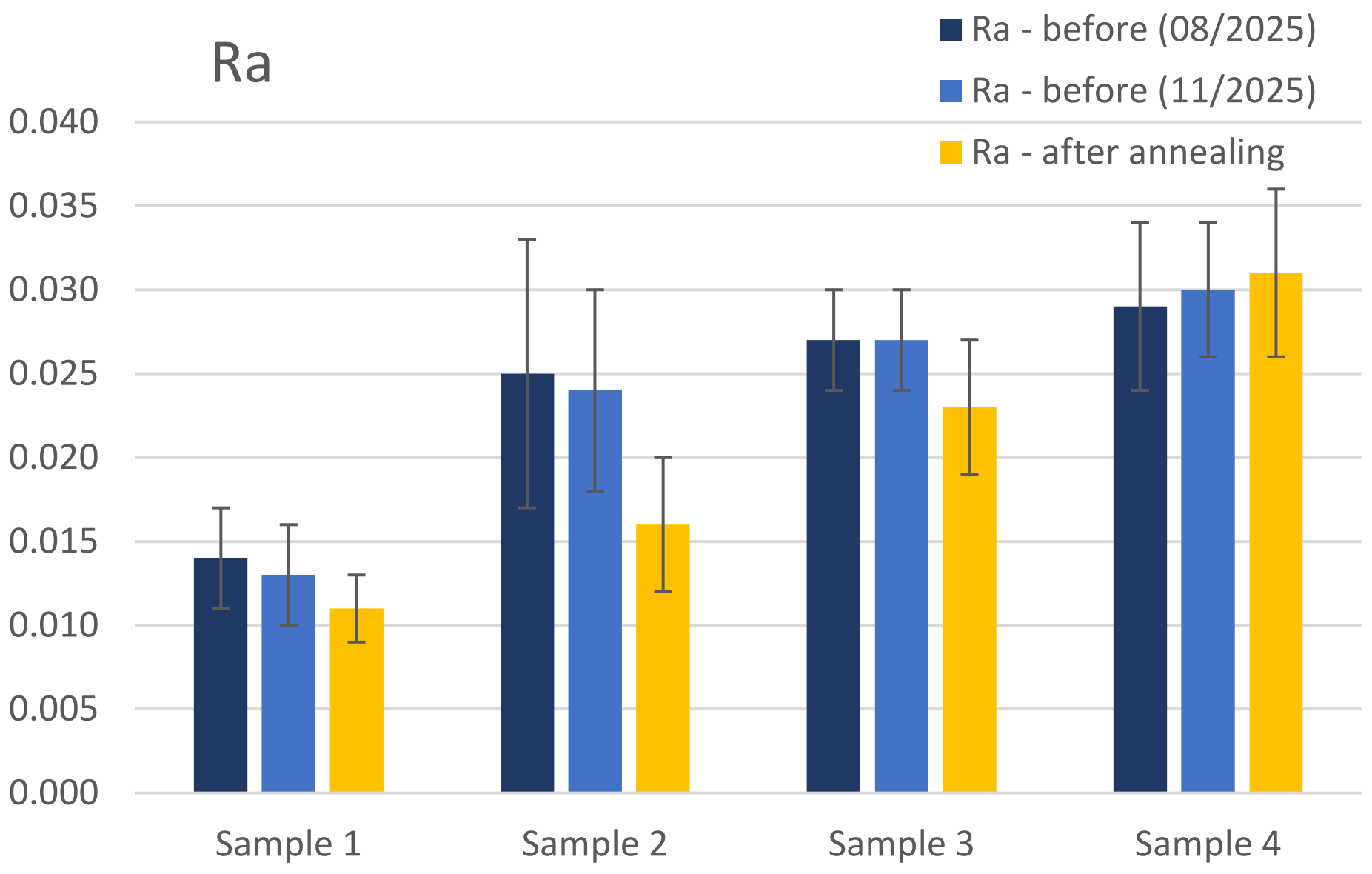
Ra
Ra - before (08/2025)
Ra - before (11/2025)
Ra - after annealing
0.040
0.035
0.030
0.025
0.020
0.015
0.010
0.005
0.000
Sample 1
Sample 2
Sample 3
Sample 4

a)

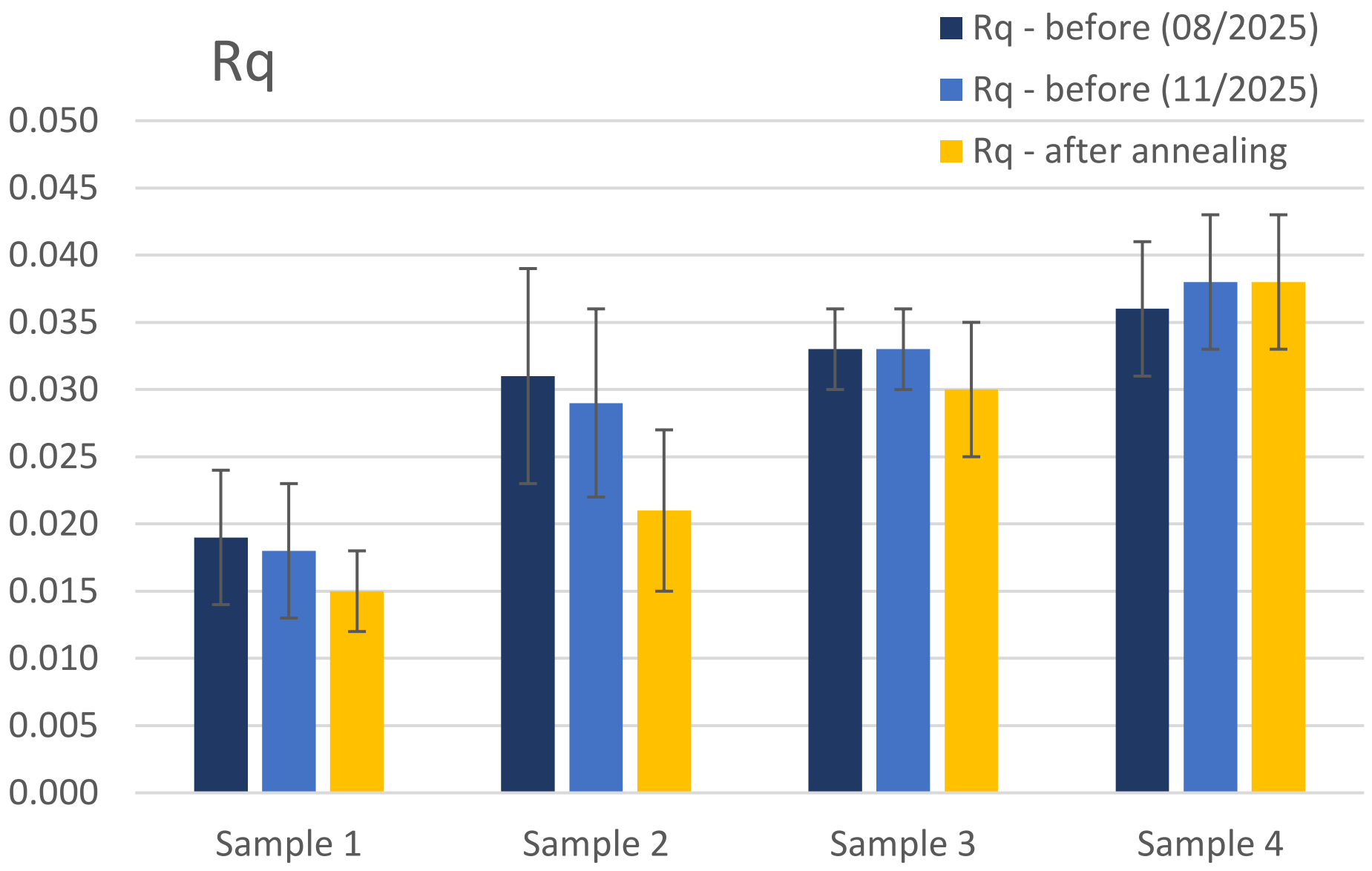
Rq
Rq - before (08/2025)
Rq - before (11/2025)
Rq - after annealing
0.050
0.045
0.040
0.035
0.030
0.025
0.020
0.015
0.010
0.005
0.000
Sample 1
Sample 2
Sample 3
Sample 4

b)

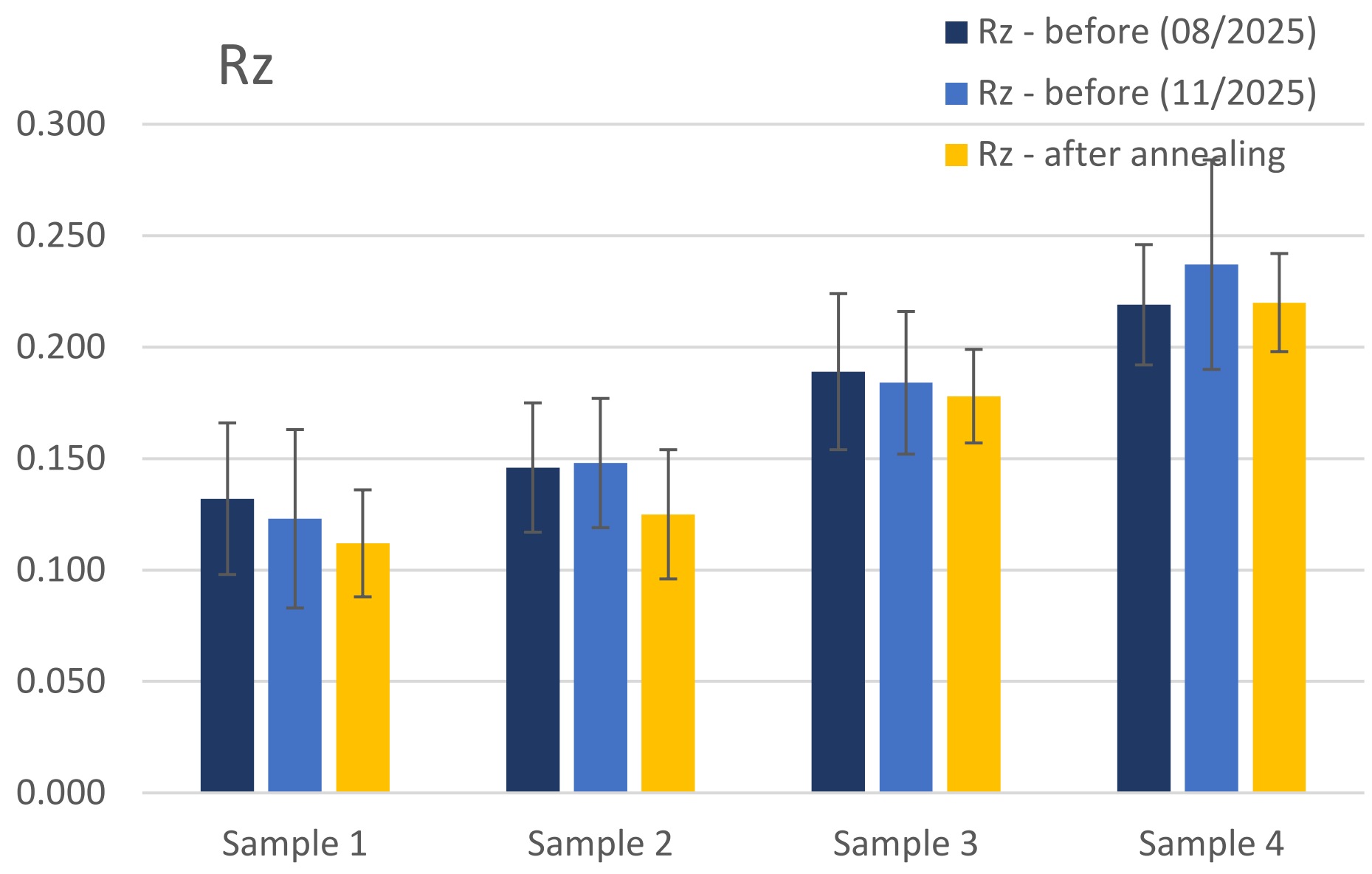

c)

Fig. 10. Roughness measurement on the free space between indents on samples 1-4 before (twice for control measurements) and after annealing

### 3.6. Grain size revelation of NiTi powder and coating by EBSD

To understand the heterogeneity behaviour of the indent recovery throughout the samples, grain size orientation was revealed within NiTi particles that were used as feedstock materials for plasma-sprayed coatings. The NiTi particle shows the phase recognized as austenite within the grains. The particles are either single-crystalline or consist of a few coarse grains several tens of micrometres in size, partly accompanied by fine-grained regions (Fig. 11). However, the plasma sprayed sample 2 (selected for its EBSD pattern quality) has chosen grain orientation in the area of residual indent marks region. The grains are distributed from fine to coarse size with various orientations within the crystalline structure of the austenite. The indent marks are visible on the grain size as black marks, which are located at various grain locations, starting from fine to coarse grain sizes. This heterogeneity of the grains may contribute to the wide distribution of indent depth recovery in the sample. The indent depth corresponds to the volume of the surface under the indent point that undergoes deformation. Since the deformed volume beneath an indent contains a different combination of fine and coarse grains at each location, this grain-size heterogeneity is a plausible contributor to the scatter in both elastic and thermal recovery, as shown on EBSD of sample 2 (see Fig.12).

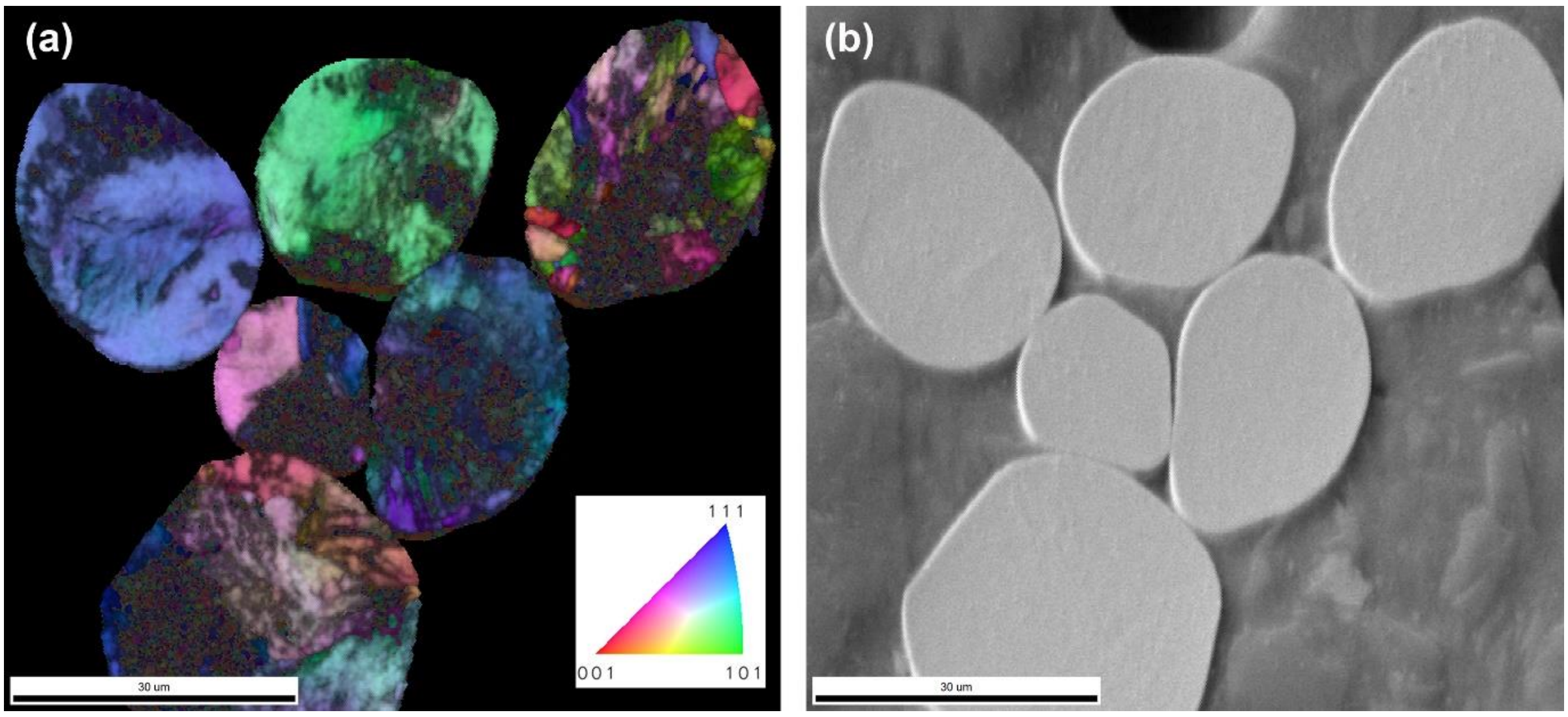

Fig. 11. The orientation map obtained by EBSD in NiTi powders: (a) the colour coded orientation map of austenitic grains (B2 phase color code is in inlay) is overlayed with gray scale of confidence index (CI) coding evaluation precision; (b) SEM image of the particles The darker areas in part (a) points to higher uncertainty of orientation evaluation in particles caused or by plastic deformation or by polishing quality.

The coatings that are generated from splats of molten particles generate the coatings of multilayers with various grain sizes. The distribution of grain sizes was earlier studied for various coating layers generated from thermal plasma-sprayed materials [21-26]. Rapid solidification of the splats produces fine grains, whereas partially melted particles can retain the coarser grains of the feedstock, resulting in a distribution from fine to coarse grains [27-29]. The plasma sprayed structure generate sub-micron to nano size grains in structural evidence [30-32].

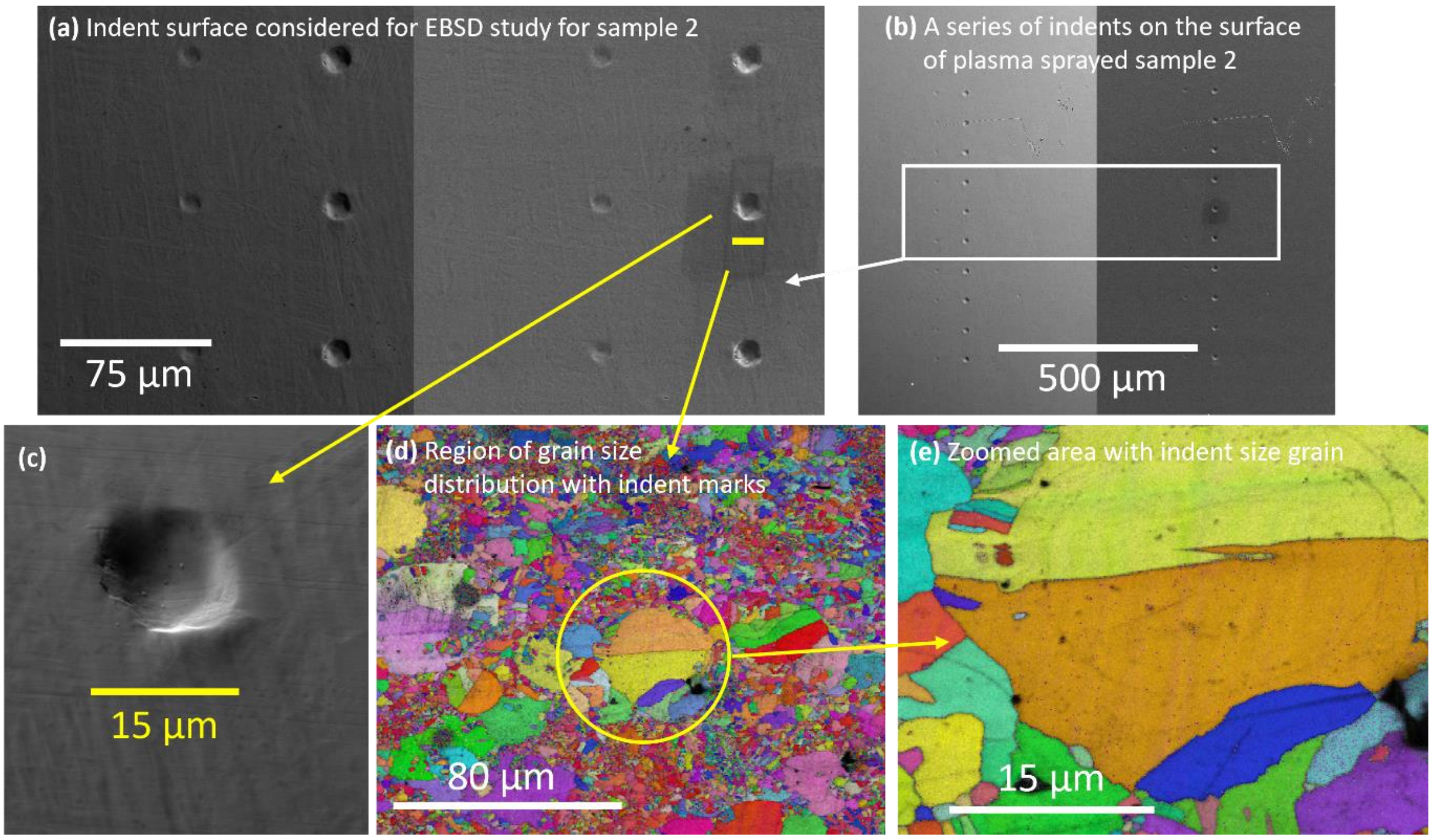

Fig. 12. The orientation maps of plasma-sprayed sample 2 (one sample is chosen out of 4) (a) Indent mark on the surface considered for this study, BSE signal on the left and SE signal on the right; the yellow bar shows average indent size, whereas white 75 μm bar shows distance between indents. (b) A series of indents across the samples, part (a) marked as white rectangle. (c) Zoomed image of one indent. (d) the orientation map overlayed with grey scale of image quality (IQ) coding the quality of EBSD patterns measured in respective points. Yellow-marked region shows correspondence with the highlighted area of various grain sizes, from finer to coarser grains, distributed (e) The enlarged area (measured with the finer step) shows the presence of various grain sizes. The color code from Fig 11a) can be used here.

## 3.7. Hardness and reduced modulus on the plasma-sprayed samples

Nanoindentation mapping of hardness and reduced modulus for samples 1–3 is presented in Fig. 13. Sample 1 is the softest and the most uniform, with hardness of 1–2 GPa and reduced modulus of 45–55 GPa, whereas Samples 2 and 3 show higher and more variable in both hardness 2–4 GPa and reduced modulus 50–70 GPa, with the largest local variations in Sample 3. The low and uniform values of Sample 1 are consistent with its high fraction of NiTi and low fraction of hard intermetallics , while the harder regions in Samples 2 and 3 are attributed to $Ni_4Ti_3$ and $Ti_2Ni$, respectively.

At first sight this ranking contradicts the spherical indentation, where Sample 2 reached the largest penetration depth at every load. The two measurements, however, probe different volumes: the mapping indents reach a depth in the range of 200–400 nm and sample individual phases at the surface, whereas the spherical indents penetrate up to 4.5 μm and engage a volume of several tens of micrometres that includes pores and splat boundaries. The larger compliance of Sample 2 under spherical loading therefore reflects the response of the deposit as a whole, including reorientation of martensite, rather than the hardness of its constituent phases.

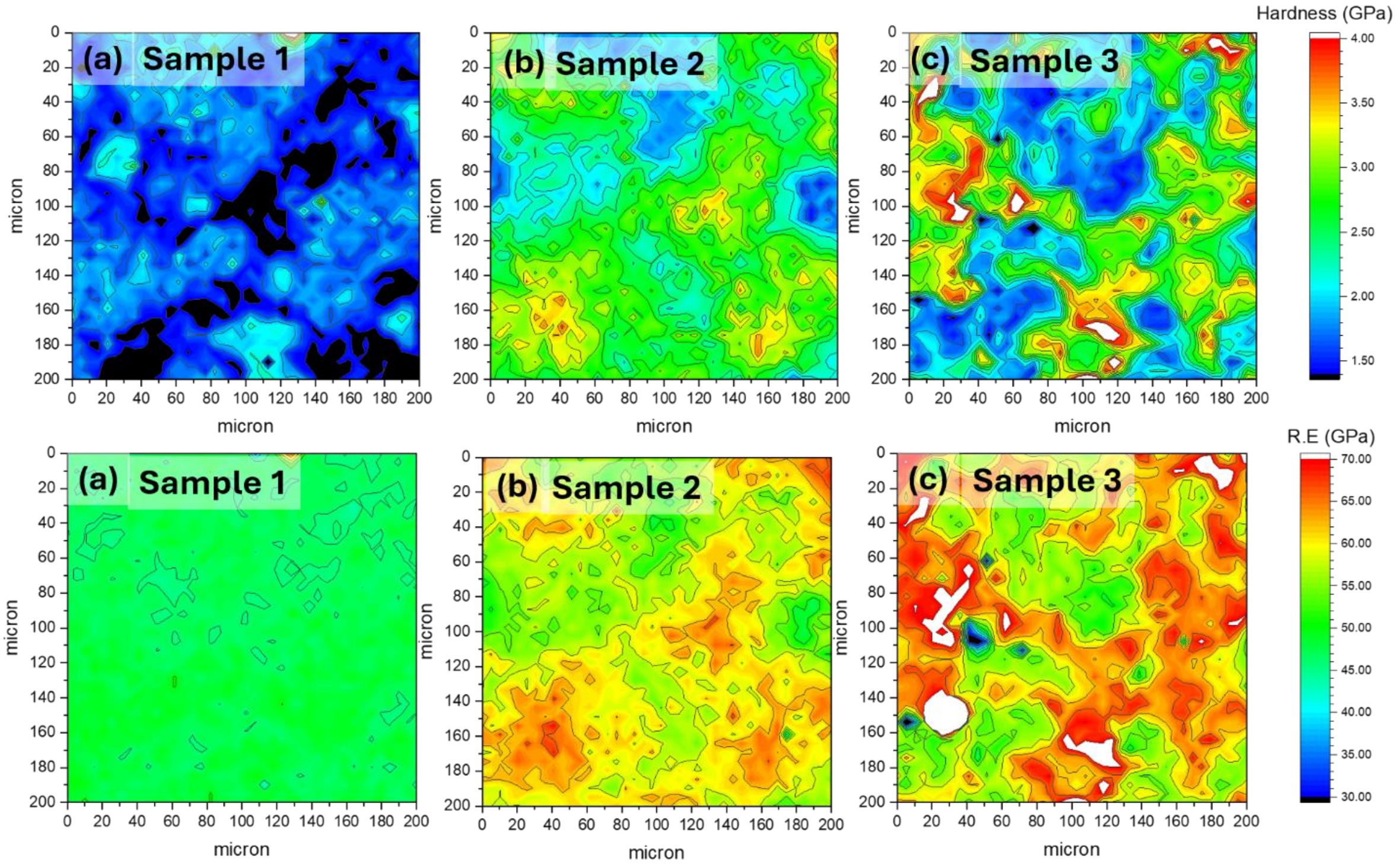

Fig. 13. Mapping of hardness and reduced modulus in samples 1-3. Sample 1 shows lower hardness (1-2 GPa) and reduced modulus (45-55 GPa), and samples 2 and 3 have hardness (2-4 GPa) and modulus (50-70 GPa).

## 3.8. Sub-micron level structural evidence

Thermal plasma sprayed samples 1, 2 were investigated by TEM microscopy. Fig. 14 (a-d) displays the bright field image and their corresponding diffraction pattern from selected area marked by red circles for sample 1 and 2.

In sample the area has been chosen for the bight field image is under 5 µm next to the spherical nano indentation marks. The top edge of the foil has been marked as bright field image. The bright field image of the sample 1 shows the single crystal with no grain boundaries visible on the image. The bending contours are visible on the bright field image of the foil in sample 1 (Fig. 14 a). The diffraction pattern of sample 1 shows austenite grains in [100] zone axis (Fig. 14 b). However, sample 2 shows twinned martensite regions with diffraction pattern confirms it (Fig. 14 c-d). The EDX analysis was carried out in a selected area of sample 1 in Fig. 15 a, that shows the presence of dislocations. The line scan in the area presents the EDX analysis, with Ni-rich NiTi matrix and presence of secondary phases such as $NiTi_2$ has been confirmed (Fig. 15 b). However, in sample 2 the matrix shows slightly richer in Ti. Moreover, the $NiTi_2$ phase shows elevated level of oxygen (approx. 5 at. %) as shown in Fig. 15 (c-d).

The Ni-rich matrix of the examined Sample 1 lamella is consistent with the retention of B2 at room temperature, since Ni enrichment lowers the transformation temperatures [15]. As each lamella samples only a few micrometres, these observations should be regarded as local.

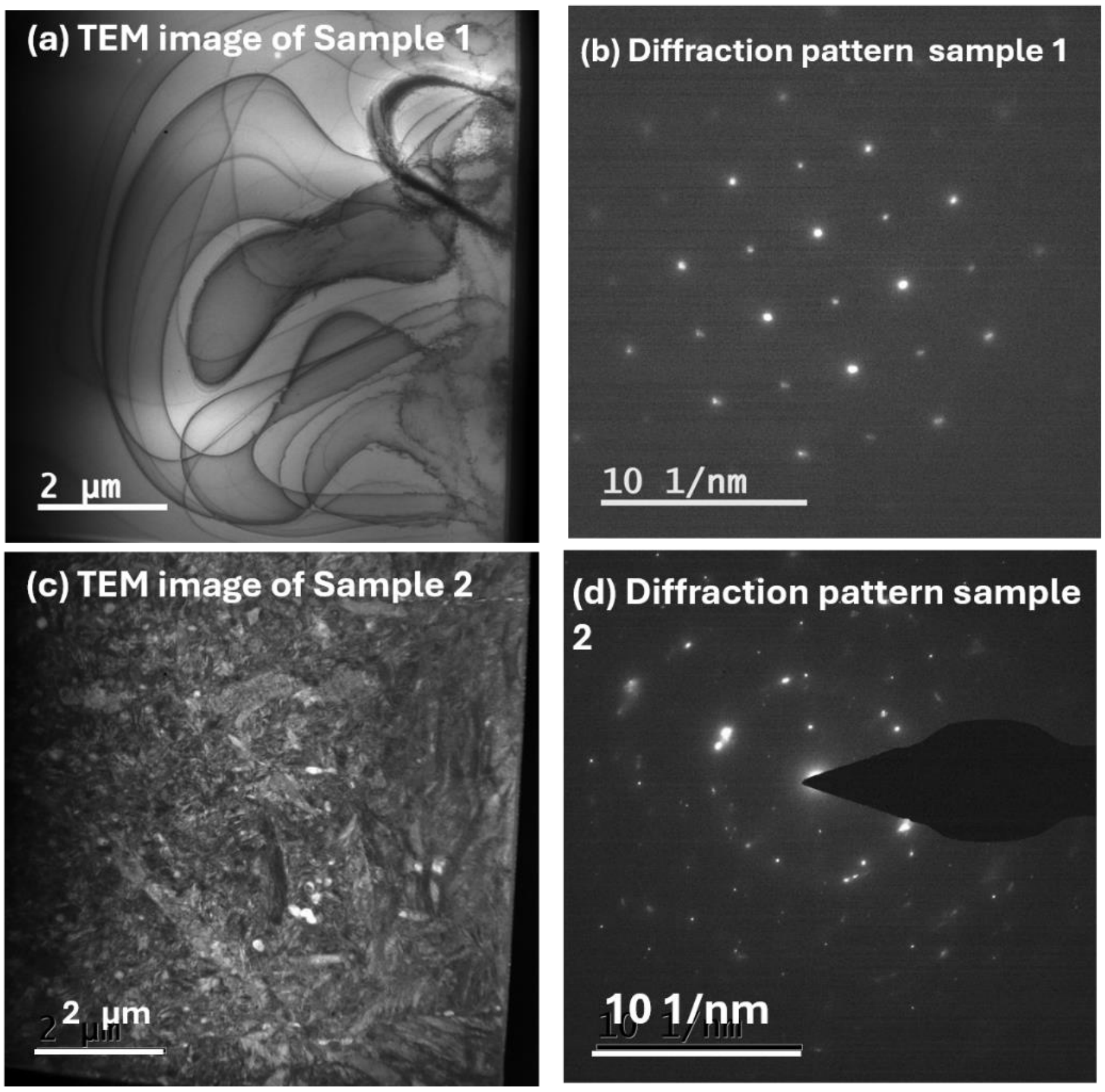

Fig. 14. (a) Bright field image of the sample 1 and (b) corresponding diffraction pattern, (c) bright field image of the sample 2 and (d) corresponding diffraction pattern.

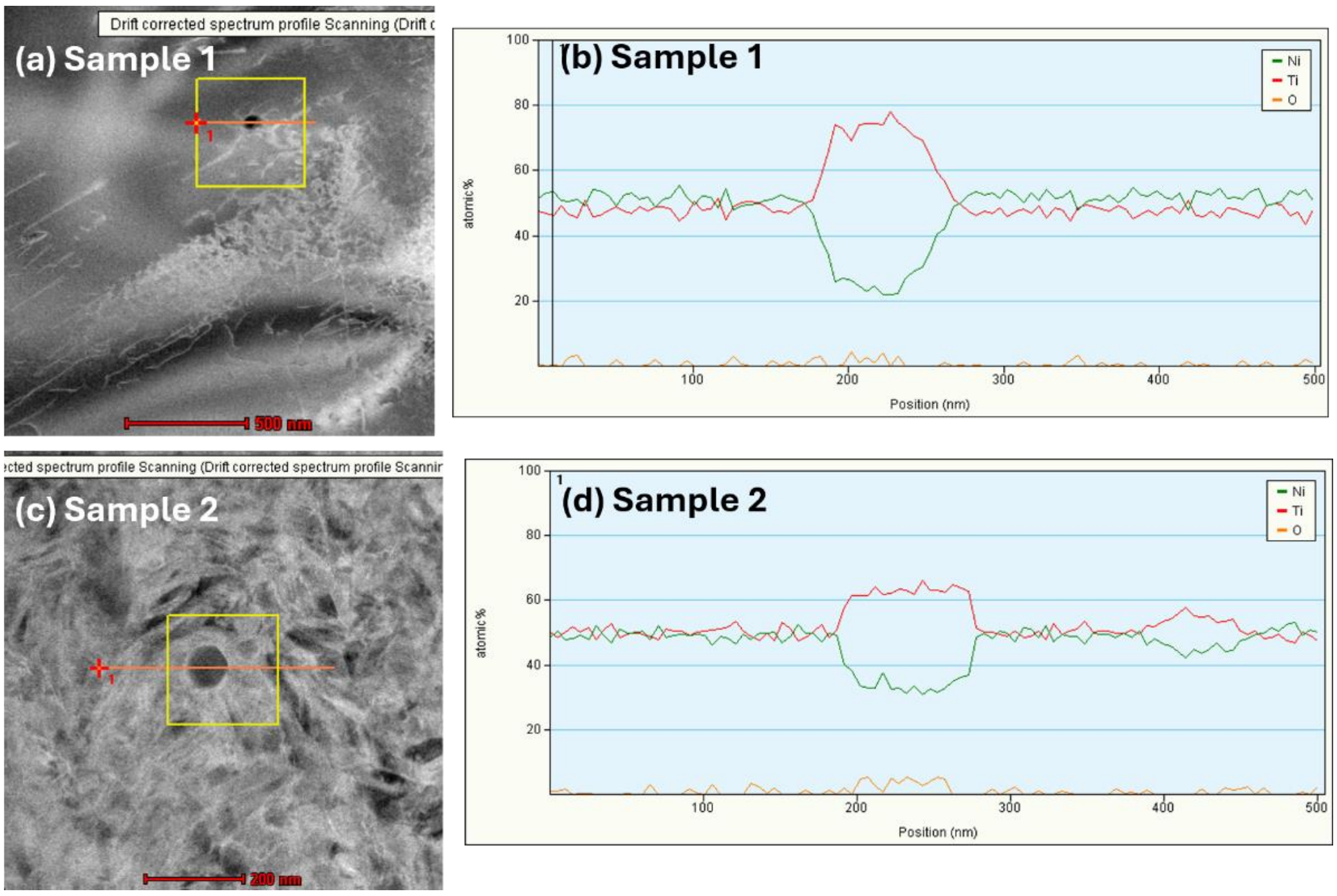

Fig. 15 EDX analysis of the sample 1 (a) and sample 2 (c) in the chosen area with line scan and the corresponding concentration graphs (b and c) showing presence of Ni, Ti and O.

## 4. Conclusions

The local mechanical response and the recovery of indentation imprints were studied in four plasma-sprayed NiTi thick coatings by spherical nanoindentation at 25–500 mN, combined with confocal profilometry before and after annealing above $A_f$. The main conclusions are:

- Nominally identical indents differ substantially in depth, most strongly at low loads, reflecting the heterogeneous splat microstructure of the coatings. In the most heterogeneous coatings, the scatter decreases with load as the deformed volume averages over an increasing number of microstructural domains.
- Both the recovery on unloading and the thermal recovery decrease with increasing load, as irreversible contributions – dislocation plasticity and pore collapse – grow faster than the recoverable transformation strain. Imprints made at the lowest loads recover essentially completely upon annealing.
- The coatings behave differently when studying elastic and thermal recovery: the coating with the largest fraction of transforming NiTi recovers least on unloading but most on heating, whereas the $Ni_4Ti_3$-containing coating shows the opposite behaviour, both regaining a comparable total fraction of the maximum depth. Since all coatings have nearly identical transformation temperatures, these differences are governed by phase composition and porosity rather than by the room-temperature phase state; the most porous deposit recovers least both elastically and thermally.

- Residual depths measured by confocal profilometry agree with those from the load–displacement curves within a few %, and annealing does not change the surface roughness, so that the disappearance of low-load imprints reflects genuine recovery.

Local indentation recovery thus provides a sensitive probe of the functional quality of plasma-sprayed NiTi and a reference dataset for powder-route NiTi surfaces.

## 5. Author Contributions

**Jan Tomáštík:** Writing – original draft, Methodology, Investigation, Data curation, Formal analysis, Conceptualization. **Sneha Samal:** Writing, Supervision, Project administration, Methodology, Investigation, Formal analysis, Data curation, Conceptualization. **Lukáš Václavek:** Writing – original draft, Resources, Methodology, Investigation, Formal analysis, Data curation. **Jaromír Kopeček**: SEM microstructure, EBSD grain size analysis, **Akhil Bhardwaj:** Nanoindentation mapping, hardness, reduced modulus analysis, Methodology, Investigation, Formal analysis, Data curation**, Tadeáš Těhan:** Porosity examination**,** Methodology, Investigation, Formal analysis, Data curation, **Ondřej Tyc**: TEM; Methodology, Investigation, Formal analysis, Data curation.

### 5.1. Acknowledgments

The authors would like to acknowledge the Institute of Plasma Physics (IPP) of the Czech Academy of Science for the plasma facilities used to prepare samples by RF-ICP method, specifically to Stanislav Habr who managed the preparation.

## 6. Funding

This research was funded by project Sensors and Detectors for Future Information Society SENDISO (No. CZ.02.01.01/00.22_008/0004596) co-funded by EU. We also acknowledge the project JAK-FerrMion (CZ.02.01.01/00/22_008/0004591) and CzechNanoLab Research Infrastructure (LM2023051) supported by MEYS CR.

## 7. Declaration of competing interest

The authors declare no conflict of interest. The funders had no role in the design of the study; in the collection, analyses, or interpretation of data; in the writing of the manuscript, or in the decision to publish the results.

## 8. Data availability

Dataset is accessible on DOI 10.5281/zenodo.24579455.

## 9. References

[1] K. Otsuka, X. Ren, Physical metallurgy of Ti–Ni-based shape memory alloys, Progress in Materials Science, 50 (2005) 511-678.

[2] S. Parvizi, S.M. Hashemi, F. Asgarinia, M. Nematollahi, M. Elahinia, Effective parameters on the final properties of NiTi-based alloys manufactured by powder metallurgy methods: A review, Progress in Materials Science, 117 (2021) 100739.

[3] M. Bram, A. Ahmad-Khanlou, A. Heckmann, B. Fuchs, H.P. Buchkremer, D. Stöver, Powder metallurgical fabrication processes for NiTi shape memory alloy parts, Materials Science and Engineering: A, 337 (2002) 254-263.

[4] S. Samal, J. Zeman, J. Kopeček, P. Šittner, Thermal plasma spraying of NiTi powder for thick coating of shape memory alloy, Journal of Alloys and Compounds, 984 (2024) 173962.

[5] S. Samal, J. Zeman, S. Habr, O. Pacherová, J. Kopeček, P. Šittner, Preparation and Characterization of Multilayer NiTi Coatings by a Thermal Plasma Process, Materials, 17 (2024) 694.

[6] B. Swain, P. Mallick, S.K. Bhuyan, S.S. Mohapatra, S.C. Mishra, A. Behera, Mechanical Properties of NiTi Plasma Spray Coating, Journal of Thermal Spray Technology, 29 (2020) 741-755.

[7] S. Samal, O. Tyc, J. Cizek, J. Klecka, F. Lukáč, O. Molnárová, E. de Prado, Z. Weiss, J. Kopeček, L. Heller, P. Šittner, T. Chráska, Fabrication of Thermal Plasma Sprayed NiTi Coatings Possessing Functional Properties, Coatings, 11 (2021) 610.

[8] S. Samal, S. Joshi, S. Björklund, M. Chandra, A. Bhardwaj, J. Kopeček, S. Habr, L. Václavek, J. Tomáštík, O. Tyc, O. Molnárová, P. Šittner, HVAF spraying of NiTi coatings: Microstructure, phase transformation and shape memory behavior, Applied Surface Science Advances, 34 (2026) 101038.

[9] S. Samal, Thermal plasma technology: The prospective future in material processing, Journal of Cleaner Production, 142 (2017) 3131-3150.

[10] Y. Lahmar-Mebdoua, A. Vardelle, P. Fauchais, D. Gobin, Heat Diffusion in Solidifying Alumina Splat Deposited on Solid Substrate under Plasma Sprayed Conditions: Application to Coating Formation, Defect and Diffusion Forum, 297-301 (2010) 46-51.

[11] Y.-H. Lu, L. Huang, M.-J. Liu, G.-J. Yang, C.-J. Li, Heterogeneous layered structure in thermal barrier coatings by plasma spray-physical vapor deposition, Journal of Advanced Ceramics, 12 (2023) 386-398.

[12] S. Samal, J. Zeman, S. Habr, O. Pacherová, M. Chandra, J. Kopeček, P. Šittner, Evaluation of Microstructure–Porosity–Hardness of Thermal Plasma-Sprayed NiTi Coating Layers, Journal of Manufacturing and Materials Processing, 7 (2023) 198.

[13] S. Samal, J. Kopeček, P. Šittner, Interfacial Adhesion of Thick NiTi Coating on Substrate Stainless Steel, Materials, 15 (2022) 8598.

[14] S. Samal, J. Tomáštík, R. Čtvrtlík, L. Václavek, O. Molnárová, P. Šittner, Surface Deformation Recovery by Thermal Annealing of Thermal Plasma Sprayed Shape Memory NiTi Alloys, Coatings, 13 (2023) 433.

[15] J. Frenzel, A. Wieczorek, I. Opahle, B. Maaß, R. Drautz, G. Eggeler, On the effect of alloy composition on martensite start temperatures and latent heats in Ni–Ti-based shape memory alloys, Acta Materialia, 90 (2015) 213-231.

[16] A. Borgaonkar, R. Bhaskaran Nair, J. Cherian Chekotu, D. Brabazon, Nanoindentation-based mapping of phase-dependent mechanical response in additively manufactured NiTi alloys, Progress in Additive Manufacturing, 11 (2026) 5145-5159.

[17] A. Jamleh, A. Sadr, N. Nomura, Y. Yahata, A. Ebihara, T. Hanawa, J. Tagami, H. Suda, Nano-indentation testing of new and fractured nickel-titanium endodontic instruments, Int. Endod. J., 45 (2012) 462-468.

[18] S. Samal, J. Tomáštík, L. Václavek, M. Chandra, J. Kopeček, I. Stachiv, P. Šittner, Recovery of deformation surface of superelastic and shape memory NiTi alloy, Applied Surface Science Advances, 25 (2025) 100684.

[19] W. Ni, Y.-T. Cheng, D.S. Grummon, Recovery of microindents in a nickel–titanium shape-memory alloy: A “self-healing” effect, Applied Physics Letters, 80 (2002) 3310-3312.

[20] W.C. Oliver, G.M. Pharr, An improved technique for determining hardness and elastic modulus using load and displacement sensing indentation experiments, Journal of Materials Research, 7 (1992) 1564-1583.

[21] P. Fauchais, G. Montavon, G. Bertrand, From Powders to Thermally Sprayed Coatings, Journal of Thermal Spray Technology, 19 (2010) 56-80.

[22] B. Kharas, G. Wei, S. Sampath, H. Zhang, Morphology and microstructure of thermal plasma sprayed silicon splats and coatings, Surface and Coatings Technology, 201 (2006) 1454-1463.

[23] K. Shinoda, M. Demura, H. Murakami, S. Kuroda, S. Sampath, Characterization of crystallographic texture in plasma-sprayed splats by electron-backscattered diffraction, Surface and Coatings Technology, 204 (2010) 3614-3618.

[24] X.-Y. Guan, T.-T. Xu, X.-K. Jiang, Y.-C. Xu, L.-C. Zhang, J.-F. Yang, Active-learning-assisted optimization and HIP induced interfacial strengthening of atmospheric plasma-sprayed ta coatings on tungsten, Surface and Coatings Technology, 539 (2026) 133959.

[25] R. Tomaszek, L. Pawlowski, L. Gengembre, J. Laureyns, Z. Znamirowski, J. Zdanowski, Microstructural characterization of plasma sprayed TiO2 functional coating with gradient of crystal grain size, Surface and Coatings Technology, 201 (2006) 45-56.

[26] S.-H. Liu, J.P. Trelles, C.-J. Li, H.-B. Guo, C.-X. Li, Numerical analysis of the plasma-induced self-shadowing effect of impinging particles and phase transformation in a novel long laminar plasma jet, J. Phys. D: Appl. Phys., 53 (2020) 375202.

[27] S. Kuroda, T.W. Clyne, The quenching stress in thermally sprayed coatings, Thin Solid Films, 200 (1991) 49-66.

[28] T. Těhan, J. Kopeček, E. Iaparova, E. Alarcon, O. Tyc, S. Samal, Mechanical-microstructural correlation on SPS-fabricated NiTi alloy with sub-micro/nano scale structural evidence, Journal of Alloys and Compounds, 1084 (2026) 191917.

[29] L. Václavek, S. Samal, J. Tomáštík, O. Molnárová, P. Šittner, Multi-cycle nanoindentation, scratch and wear test on super-elastic (SE) and shape memory (SM) NiTi alloy, Results in Surfaces and Interfaces, 23 (2026) 100768.

[30] S. Samal, D. Vokoun, I. Stachiv, P. Šittner, Simulation of thermal stress predictions for NiTi coating on a stainless-steel substrate using thermal plasma spraying, Ceram. Int., 51 (2025) 12337-12345.

[31] M. Khan, Y. Zeng, Microstructural Evolution and Thermal Transport in APS SrZrO3 Coatings: An EBSD-Focused Study, Coatings, 16 (2026) 729.

[32] H. Zhang, X. Zhang, J. Zhang, M. Pan, Vacuum plasma sprayed AlCoCrFeNi high-entropy alloy bond Coating: Vacuum heat-treatment, microstructure evolution and oxidation behavior analysis, Intermetallics, 184 (2025) 108836.